\documentclass[12pt,preprint]{aastex}

\bibdata{refs}
\bibstyle{apj}

\newcommand{\ppp}{{\it ppp}}

\def\bt{{\mbox{\boldmath $\theta$}}}
\def\bs{{\mbox{\boldmath $\sigma$}}}

\shorttitle{Line Searches in {\it Swift\/} X-ray Spectra}
\shortauthors{Hurkett et al.}

\begin{document}


\title{Line Searches in {\it Swift\/} X-ray Spectra}


\author{C. P. Hurkett$^{1}$\altaffilmark{*}, S. Vaughan$^{1}$, J. P. Osborne$^{1}$, P. T. O'Brien$^{1}$, K. L. Page$^{1}$, A. Beardmore$^{1}$, O. Godet$^{1}$, D. N. Burrows$^{2}$, M. Capalbi$^{3}$, P. Evans$^{1}$, N. Gehrels$^{4}$, M. R. Goad$^{1}$, J. E. Hill$^{4,5}$, J. Kennea$^{2}$, T. Mineo$^{6}$, M. Perri$^{3}$ \& R. Starling$^{1}$}

\affil{$^{1}$XROA Group, Dept. of Physics and Astronomy, University of Leicester, Leicester LE1 7RH, UK. \\ $^{2}$Department of Astronomy and Astrophysics, Pennsylvania State University, USA. \\ $^{3}$ASI Science Data Center, ASDC c/o ESA-ESRIN, via G. Galilei 00044 Frascati, Italy. \\ $^{4}$NASA Goddard Space Flight Center, Greenbelt, MD 20771, USA. \\ $^{5}$Universities Space Research Association, 10211 Wincopin Circle, Suite 500 Columbia, MD 21044, USA. \\ $^{6}$INAF IASF-Pa, via U. La Malfa 153, 90146 Palermo, Italy.}


\altaffiltext{*}{cph9@star.le.ac.uk}


\begin{abstract}
Prior to the launch of the {\it Swift\/} mission several X-ray line detections were reported in Gamma Ray Burst 
afterglow spectra. To date, these pre-{\it Swift\/} era results have not been conclusively confirmed. 
The most contentious issue in this area is the choice of statistical method used to evaluate the significance of these features. 
In this paper we compare three different methods already extant in the literature for assessing the significance
of possible line features and discuss their relative advantages and disadvantages. The methods are demonstrated by application to 
observations of 40 bursts from the archive of {\it Swift\/} XRT at
early times ($<$ few ks post trigger in the rest frame of the
burst). Based on this thorough analysis we found no strong evidence for emission lines.
For each of the three methods we have 
determined detection limits for emission line strengths in bursts with
spectral parameters typical of the {\it Swift\/}-era sample. 
We also discuss the effects of the current calibration status on emission line detection.
\end{abstract}





\section{Introduction} \label{section:Intro}

It is widely accepted that the spectra of the X-ray afterglow of Gamma-Ray Bursts (GRBs) are dominated by non-thermal emission, the leading 
candidate for which is synchrotron emission (\citealt{Piran_2005} and references therein), though alternate emission processes have also been 
suggested such as self-Compton (\citealt{Waxman_1997} and \citealt{Ghisellini_1999}) or inverse Compton scattering of external light (\citealt{Brainerd_1994};
\citealt{Shemi_1994}; \citealt{Shaviv_1995}; \citealt{Lazzati_2004}).

Up to the present time the X-ray spectra of \textit{Swift} afterglows are generally well described by an absorbed power law (for counter-examples 
see \citealt{Butler_2007}), typically absorbed by material with a column density in excess of the well measured Galactic values~\citep{Campana_2006_AA}. 
Table 2 of \citet{Campana_2006_AA} shows that, of the 17 bursts analyzed, 14 have observed $N_{\rm H}$ values greater than the measured Galactic column 
density, whilst the remaining three have observed $N_{\rm H}$ values that are consistent, within limits, with the measured values.

In the past it has been proposed that there are other spectral features, with varying levels of significance, in addition to the basic absorbed 
power law spectrum (\citealt{Piro_1999}; \citealt{Yoshida_1999}; \citealt{Amati_2000} \citealt{Antonelli_2000}; \citealt{Piro_2000}; \citealt{Reeves_2002};
\citealt{Watson_2002}; \citealt{Watson_2003} and \citealt{Frontera_2004}). Most are attributed to Fe $K_{\alpha}$ emission lines or the radiative 
recombination continuum of the same element. Some have been attributed to the $K_{\alpha}$ lines of Ni, Co or of lighter elements such as Si, S, 
Ar and Ca. In two cases there has been a report of a transient absorption feature also corresponding to Fe $K_{\alpha}$ (\citealt{Amati_2000}; 
\citealt{Frontera_2004}).

The models for the production of such emission features are divided into transmission and reflection models, though the large equivalent widths 
($\sim$ few keV) inferred from the observed X-ray features favor models in which the line is produced by reflection (\citealt{Rees_2000}; 
\citealt{Ballantyne_2001} and \citealt{Vietri_2001}). Proposed models have to overcome two constraints; the {\it size problem\/} and the 
{\it kinematic problem\/}. Observing a line at a time $t_{obs}$ after the burst implies that the emitting material must be within a distance of 
$\sim ct_{obs}/(1+z)$ from the central engine, thus implying that the region must be compact if a line, or lines, are observed at early times. 
Additionally the emitting region must contain $\sim0.1 M_{\odot}$ of Fe (in the case of Fe K$_{\alpha}$ features) whilst still being optically 
thin to electron scattering, in order that Comptonization does not broaden the line beyond the observed widths~\citep{Vietri_2001}. If the line 
width is interpreted as being due to the velocity of the supernova remnant, the observed limit on this width implies an age limit on the remnant 
of $\sim10-20$ days. However, at this time, Co nuclei outnumber both Ni and Fe nuclei; thus the emission line would be due to Co at an energy of 
$7.5/(1+z)$ keV, which is the {\it kinematic problem\/}.

Various geometries have been suggested for the reflection models, which rely on either a precursor or simultaneous supernova (SN) event. If a SN occurs
several tens of days before the GRB this solves both the size and kinematic problems. In these cases the radiation from the GRB jets can either illuminate 
the inner face of the SN shell remnant or the inner faces of wide
funnels that they excavate through young plerionic remnants. However,
these models have been 
questioned following the simultaneous GRB-SN association indicated by
GRB 980425~\citep{Galama_1998} and then confirmed by GRB 030329 
(\citealt{Hjorth_2003}; \citealt{Stanek_2003}) and  GRB 060218
(\citealt{Campana_2006}; \citealt{Pian_2006}). In this case the most
likely scenario for emission  
line production occurs if the progenitor ejects a large amount of matter, at subrelativistic speeds, along its equator. The halo of material surrounding 
massive stars, ejected by their strong stellar winds towards the end
of their main sequence lifetime, scatters a fraction of the photons
from the prompt and  
afterglow phase back into the equatorial material, which then produces X-ray line emission \citep{Vietri_2001}. 

Verifying the presence of such spectral features is of critical
importance as they will allow us to probe the circumburst environment
of the GRB as well  
as gaining an indirect indication of the possible structure and
behavior of the central engine. 

The statistical significance of the 1999-2003 reported features is low
(usually $2-3\sigma$), only two detections have a significance
$>$4$\sigma$  (GRB 991216: 4.7$\sigma$, \citealt{Piro_2000}; and GRB
030227: 4.4$\sigma$, \citealt{Watson_2003}). Though later detections
were made with much more sensitive  
instruments than the early ones, all detections have remained at this
low significance level, as a result they remain the subject of
much debate. Arguably the most controversial issue in the
discussion of line detections is the choice of statistical method employed to
gauge their significance. At least four
methods have been proposed and used in the GRB literature:
\begin{enumerate}

\item The likelihood ratio test (LRT) and related $F$-test

\item Bayes factors

\item Bayesian posterior predictive probability

\item Monte Carlo test for peaks in data after `matched filter' smoothing

\end{enumerate}
Examples of the application of these methods to GRB X-ray spectra can be found in: \citet{Freeman_1999}, 
\citet{Yoshida_1999}, \citet{Piro_1999}, \citet{Protassov_2002}, \citet{Rutledge_2003}, \citet{Tavecchio_2004}, 
\citet{Butler_2005}, \citet{Sako_2005}, \citet{Butler_2007} and references therein.  

In all the applications cited above, an underlying continuum was
assumed, usually in the form of an absorbed power law (e.g. using the
Wisconsin absorption model, see \citealt{Morrison_1983}). 
The detection of a line then 
amounts to a comparison of two models: $M_0$, the simple ``continuum''
model, and
$M_1$, the more complex ``continuum $+$ line'' model. 
The strength, location and width of the emission line may be
restricted or allowed to be free parameters.

As discussed in depth by \citet{Protassov_2002} and
\citet{Freeman_1999}, there are strong theoretical reasons why
the LRT is not suitable for assessing the significance of
emissions lines, despite its popularity in the literature.
We will not repeat those arguments here. 
It is the purpose of the present paper to
compare the relative merits of the remaining three methods, in terms
of their computational efficiency, robustness and sensitivity limits by
applying all three methods to X-ray spectra from the {\it Swift\/} archive.
This is a particularly rich archive because 
of the combination of the rapid slew response of the {\it Swift\/} GRB mission
\citep{Gehrels_2004} and the powerful X-ray Telescope~\citep[XRT;][]{Burrows_2005}.

The remainder of this paper is organized as follows. In $\S$\ref{section:Data_reduction} we provide 
details of the sample selection criteria and basic data reduction. The theoretical basis and practical 
applications of the three statistical methods under investigation are described in detail in 
$\S$\ref{section:Analysis_methods}. In $\S$\ref{section:PKS} we apply all of the methods to PKS 0745-19, 
a known line emitting source, to demonstrate the expected outcome when a line is present, and 
$\S$\ref{section:Detection_limits} discusses a simulation study to assess the line detection limits of 
each method for typical \textit{Swift} XRT data. In $\S$\ref{section:Swift} we discuss the results from 
the \textit{Swift} archival GRB afterglow data, highlighting several GRBs with potential additional spectral 
components. $\S$\ref{section:disco} is dedicated to a discussion of our results and their comparison to other 
recent line detections in the literature. Finally, $\S$\ref{section:conc} presents our conclusions.

\section{Data Reduction} \label{section:Data_reduction}

This paper reports on the analysis of Windowed Timing (WT) mode data from GRB 050128 to GRB 060510B, covering a total of 153 bursts, 40 of which 
contained sufficient WT mode data for our analysis methods. WT mode data was chosen primarily because the time interval covered by these observations, 
typically T+0 s to T+~500 s (though for bright bursts this may extend up to T+few ks) in the rest frame of the burst, is rarely explored. 
Prior observations of the 1999-2003 bursts typically start at 20+ hours after the trigger in the observer's reference frame, though
\citet{Antonelli_2000} report on an emission line detection at T+$\sim 12$ hours; \citet{Amati_2000} and \citet{Frontera_2004} report absorption line 
features in very early time data (T+$<$20 s and T+$<$300 s respectively) from the WFC of {\it  BeppoSAX\/}. Additionally WT mode data is only taken whilst 
the GRB afterglow is bright.  
All of the methods discussed can easily be extended to Photon Counting
(PC) mode data. We acknowledge that the current theoretical models for
line emission  
indicate that lines could occur at times not covered by WT mode data, however, the same models do not rule out this time period either.  

All data have been obtained from the UK {\it Swift\/} archive\footnote{http://www.swift.ac.uk/swift$\_$live/obscatpage.php}~\citep{Tyler_2006}
and processed through {\it xrtpipeline} v0.10.3\footnote{Release date 2006-03-16} using version 008 calibration files and correcting for 
the WT mode gain offset (if present). Version 008 of the CALDB is a marked improvement over the previous release
~\citep{Campana_2006_caldb}\footnote{http://heasarc.gsfc.nasa.gov/docs/heasarc/caldb/swift/docs/xrt/SWIFT-XRT-CALDB-09.pdf}, however, it may be the case 
that low energy calibration features have still not been optimally corrected. We did not apply any systematic correction factor to the errors of our
spectra because the recommended factor is very much smaller than the statistical errors in our spectra. Grade $0-2$ data, using extraction regions of 
20$\times$3 pixels, for both source and background regions have been used. At count rates below 
100 counts s$^{-1}$ WT mode data does not suffer from
pile-up~\citep{Romano_2006}, however, some of the time intervals
considered contain sufficient flux to  
cause pile-up effects. Following \citet{Romano_2006} we have excluded central regions when necessary as detailed in their Appendix A, splitting the 
20$\times$3 pixel region into two 10$\times$3 pixel regions placed either side of the central exclusion region.

All spectral fitting and simulations have been carried out using {\it XSPEC\/} version 12.2.1ab or higher with background subtracted spectra binned 
to $\ge$20 counts bin$^{-1}$. This binning permits the use of the
$\chi^2$ minimization as a Maximum Likelihood method.
Data for each GRB being considered have been time-sliced with the following criteria in mind:
\begin{enumerate}

\item Each spectrum must contain $800-1600$ background
  subtracted counts. This is a compromise between good time resolution
  and spectral quality.

\item If one or more flares are present in the data, wherever possible
  (not violating condition 1), separate spectra were extracted for the
  rising and falling sections of the flare, since spectral evolution
  is likely to occur at this time (\citealt{Burrows_2005Sci},
  \citealt{Zhang_2007} and references therein). 

\item If data are affected by pile-up then these time periods were
  extracted separately from the non piled-up data. 

\end{enumerate} 

The range of $800-1600$ counts was chosen to ensure good time resolution while maintaining sufficient
counts to obtain a reasonable spectrum, with $\sim 40-80$ spectral bins (each with $\sim 20$ counts)
over the useful bandpass. The data considered here were taken during the early, bright phases of the
afterglow evolution (typically T+0 s to T+500 s), during which the X-ray flux and (possibly) spectrum
are often highly variable. Time resolution is therefore important to reduce the effects of
flux/spectral variation on the modelling of individual spectra. Furthermore, previous claims of
emissions lines have often reported the features as transient and so time resolution may be important
for detecting lines.

\section{Analysis Methods} \label{section:Analysis_methods}

As noted in $\S$\ref{section:Intro} several different methods have been used in the past to assess the significance of line detections in the X-ray 
spectra of GRBs. The three methods that are the subject of the present
paper are discussed individually in the following subsections.
The reader interested only in the application of these methods may
wish to skip to \S~\ref{section:PKS}.

\subsection{Bayes Factors} 
\label{subsection:Bayes_analysis_method}

The goal of scientific inference is to draw conclusions about the
plausibility of some hypothesis or model, $M$, based on the available
data $\mathbf{D} = \{ x_1, x_2, \ldots, x_N \}$, given the background information $I$
(such as the detector calibration, statistical distribution of the
data, etc.). However, when
presented with data it is usually not possible to compute this
directly. What can be calculated directly in many cases is the sampling distribution for
data assuming the model to be true, $p(\mathbf{D}|M,I)$. This is
usually called the {\it likelihood} when considered as a function of
$M$ for fixed $\mathbf{D}$. Statements about data
conditional on the model may be related to statements about the model conditional on
the data by Bayes' Theorem\footnote{For general references relating to
Bayesian analysis see http://www.astro.cornell.edu/staff/loredo/bayes/}.
In its usual form Bayes theorem relates the likelihood to the {\it posterior probability\/} of the model
$M$ conditional on data $\mathbf{D}$ (and any relevant background information
$I$), written $p(M|\mathbf{D},I)$:
\begin{equation}
p(M|\mathbf{D},I) = \frac{p(\mathbf{D}|M,I)p(M|I)}{p(\mathbf{D}|I)}.
\label{equation:bayes_eqn}
\end{equation}
The term $p(M|I)$ is the {\it prior probability\/} of the model $M$ and
describes our knowledge (or ignorance) of the model prior to
consideration of the data (often called simply the `prior'). 
The term $p(\mathbf{D}|I)$ is effectively a normalization term and is
known as the {\it prior predictive probability\/} (it describes the
probability with which one would predict the data given only prior
information about the model). For a more general discussion of Bayes 
theorem see \citet{Lee_1989}, \citet{Loredo_1990}, \citet{Loredo_1992}, \citet{Sivia_1996}, \citet{gelman},
\citet{gregory2005} and for discussion in the context of GRB line
searches see \citet[their $\S$3.1.2]{Freeman_1999} and
\citet{Protassov_2002}. In the rest of this paper we drop the explicit
conditioning on background information $I$, but it is taken as
accepted that ``no probability judgements can be made in a vacuum''
\citep{gelman}. 
 
One simple way to represent the posterior probabilities for two
alternative models is in terms of their ratio, the {\it posterior odds} (see \citealt{gregory2005} $\S$ 3.5).
This eliminates the
$p(\mathbf{D})$  term (which has no dependence on $M$). 
If we define two competing models, such as one with a line ($M_{1}$) and one
without ($M_{0}$), we
may compute the posterior odds:
\begin{equation}
  O_{10} = \frac{p(M_{1}|\mathbf{D})}{p(M_{0}|\mathbf{D})} = 
  \frac{p(M_{1})}{p(M_{0})}
  \frac{p(\mathbf{D}|M_{1})}{p(\mathbf{D}|M_{0})} =
  \frac{p(M_{1})}{p(M_{0})} B_{10}.
\label{equation:odds_1}
\end{equation}
High odds indicate good evidence for the existence of a line in the spectrum. The first term on the right hand side
is the ratio of the priors, the second term is the ratio of the
likelihoods and is often called the {\it Bayes factor} (see
\citet{Kass} for a detailed review).
In the present context we have no strong theoretical grounds to prefer
one or the other model (line or no line) and so assign equal prior probabilities to our two models.
Thus the ratio of the priors in equation~\ref{equation:odds_1} is set
to unity and the posterior odds are equal to the Bayes factor. In the
following we use the terms posterior odds, odds and Bayes factors interchangeably.

The likelihood functions in equation~\ref{equation:odds_1} are
functions of $M_i$ only. If the models contain no free parameters
(i.e. are completely specified) then equation~\ref{equation:odds_1}
can be used directly.
However, if the model does contain free parameters, the likelihood
will be a function of the parameter values.
In the present context, where the particular values of the parameters
are not the subject of the investigation, the parameters are referred
to as {\it nuisance} parameters. 
In order to remove the dependence on these nuisance parameters 
the likelihood function must be written as a function of the $N$ parameters
(denoted $\bt = \{ \theta_1,\theta_2, \ldots , \theta_{N} \}$) and integrated, or {\it
marginalized\/}, over  
the prior probability density function (PDF) for the parameters: 
\begin{equation}
  p(\mathbf{D}|M) = \int p(\mathbf{D}, \bt | M) d\bt = 
  \int p(\mathbf{D} | \bt, M) p(\bt | M) d\bt.
\label{equation:margin}
\end{equation}
The marginal likelihood is obtained by integrating over all parameter
values the joint PDF for the data and the parameters. This joint
PDF may be separated into the product of two terms using the rules of probability theory:
$p(\mathbf{D} | \bt, M)$ is the likelihood function of the data as a
function of the model and its parameters, and $p(\bt | M)$ is the prior
for the model parameters. Once these are assigned one can
compute the necessary likelihood (a function of $M$ alone) by integration.
The Bayes factor for model $M_1$ (with parameters $\bt_1$) against
model $M_0$ (with parameters $\bt_0$) may now be written:
\begin{equation}
B_{10} = \frac{ \int p(\mathbf{D} | \bt_1, M_1) p(\bt_1 | M_1) d\bt_1}{\int p(\mathbf{D} | \bt_0, M_0) p(\bt_0 | M_0) d\bt_0}.
\label{equation:bfactor}
\end{equation}
The issues of how the relevant likelihoods and priors are assigned, and
the integrals computed, are discussed below.

\subsubsection{Application to high count X-ray spectra}
\label{section:counts}

In the limit of a large number of counts per spectral
bin, the Poisson distribution of counts in each bin
will converge to the Gaussian distribution, and in this 
case equation~\ref{equation:margin} can be written in terms of
the familiar $\chi^2$ fit statistic \citep{eadie}:
\begin{equation}
\label{equation:logl}
  L = 
  \ln [ p(\mathbf{D}|\bt, M)] = 
  -\frac{1}{2}\sum_{i=1}^{N} \ln[2 \pi \sigma_i^{2}] - \sum_{i=1}^{N}
  \frac{(x_i - \mu_i(\bt))^2}{2 \sigma_i^2} =
  {\rm constant}- \chi^2/2 
\end{equation}
where $\sigma_i$ is the error on the data (e.g. counts) measured in the $i$th channel,
and $\mu(\bt)_i$ is the predicted (e.g. model counts) in the channel
based on the model with parameter values $\bt$.
The last equality can be made since the term $\sum \ln[2 \pi
\sigma_i^{2}]$ is constant given data with errors $\sigma_i$.
This is why, in the high count limit, finding the parameter values at
which $\chi^2$ is minimized is equivalent to
finding the Maximum Likelihood Estimates (MLE) of the parameters, $\hat{\bt}$.

\subsubsection{Approximating the posterior}
\label{section:applic}

In general the integrals of equation~\ref{equation:margin} and
\ref{equation:bfactor} must be
computed numerically, using for example Markov Chain Monte Carlo
(MCMC) methods (\citealt{gelman} chapter 11; \citealt{gregory2005}
chapter 12), which is computationally demanding.  However, maximum
likelihood theory says that the MLE will become more Gaussian and of
smaller variance as the sample size (number of counts)
increases, even if the model is non-linear (chapter 11 of \citealt{gregory2005}).
Therefore, with sufficient counts the likelihood function
will approach a multidimensional Gaussian with a peak at the MLE
location $\hat{\bt}$, i.e. the location of the best fit (minimum
$\chi^2$) in parameter space. Furthermore, if the prior function is relatively flat around the peak of the
Gaussian likelihood we may approximate the prior term in equation~\ref{equation:margin} by a 
constant, namely its value at the best fit location, $p(\hat{\bt} | M)$.
Putting this together means we may approximate the posterior as a Gaussian -- often called  the
\emph{Laplace approximation} -- which greatly
simplifies the integrations in equations~\ref{equation:margin}
and \ref{equation:bfactor}, since a multidimensional Gaussian may be
evaluated analytically, once its peak locations and covariances
are known, which avoids the need for computationally expensive numerical integration.

The integral of an unnormalized multidimensional Gaussian is
$(2\pi)^{N/2} \sqrt{\det[\bs^2]}$ times the peak value, where $\bs^2$
is the covariance matrix\footnote{
\label{fn:cov}
The covariance matrix is the square,
symmetric matrix comprising the covariances of parameters $\theta_i$
and $\theta_j$ as element $\sigma_{ij}^2$. By symmetry
$\sigma_{ij}^2=\sigma_{ji}^2$. The diagonal elements are the variances of
the parameters. The covariance matrix may be estimated as minus the inverse
the Hessian matrix listing all the second derivatives of the log likelihood
function $[\nabla \nabla L]_{ij} = \partial^2 L / \partial
\theta_i \partial \theta_j$. Given that $L = \log [ p(\mathbf{D}|\bt,M)] = {\rm constant} - \chi^2/2$ (in the limit of
many counts per channel) the covariance matrix may be 
estimated using $[\sigma^2]_{ij} = 2 [ ( \nabla \nabla \chi^2 )^{-1} ]_{ij}$.
The second derivatives of the $\chi^2(\bt)$ function can be evaluated numerically.
}
evaluated at the peak (best fit location) and
$N$ is the number of parameters. We may now re-write
equation~\ref{equation:margin} as: 
\begin{equation}
p(\mathbf{D}|M_i) = \exp(-\chi_{(i)}^2/2) (2 \pi)^{N_i / 2}
\sqrt{ \det[\bs_{i}^{2}]} p(\hat{\bt_i} | M_i),
\end{equation}
which involves the prior density only at the mode of the $p(\hat{\bt_i} | M_i)$
likelihood (i.e. the density at the MLE position) for each model. This can be substituted into
equation~\ref{equation:odds_1} to give the Bayes factor (see
\citealt{gregory2005} chapters 10--11):
\begin{equation}
B_{10} = \exp(- \Delta \chi^2/2)
(2 \pi)^{\Delta N / 2}
\frac{\sqrt{ \det[\bs_{1}^{2}]}}{\sqrt{ \det[\bs_{0}^{2}]}}
\frac{p(\hat{\bt}_1|M_1)}{p(\hat{\bt}_0|M_0)},
\label{equation:odds_2}
\end{equation}
\noindent where $\Delta N = N_{1}-N_{0}$ and $\Delta \chi^2 = \chi_{1}^2 - \chi_{0}^2$. This can be calculated, using the appropriate
values of $\chi^{2}$ and the covariance matrix $\bs^{2}$ evaluated at the best
fit location for each model, once we assign prior densities to each parameter.
See \citet{gregory2005}, chapters 10--11 for a discussion of
essentially the same method.

\subsubsection{Validity of the Laplace approximation}
\label{sect:approx}

There are a number of ways to check the validity of this assumption.
One is to inspect the shape of the $\chi^2(\bt)$ surface, which is related
to the likelihood surface by  $L \sim -\chi^2/2$. A Gaussian
likelihood is equivalent to a paraboloidal log-likelihood or
$\chi^2(\bt)$. If the contours of $\Delta \chi^2$ appear paraboloidal
around the mimimum,
in one and two dimensions, for each parameter or pair of parameters,
the likelihood surface must be approximately Gaussian (the conditional
and marginal distributions of a Gaussian are also Gaussian, so slices
through the $\chi^2$ space should also be paraboloidal).
This was generally true for the continuum model for the XRT data.

\begin{figure}
\begin{center}
     \includegraphics[width=5.8cm,angle=90]{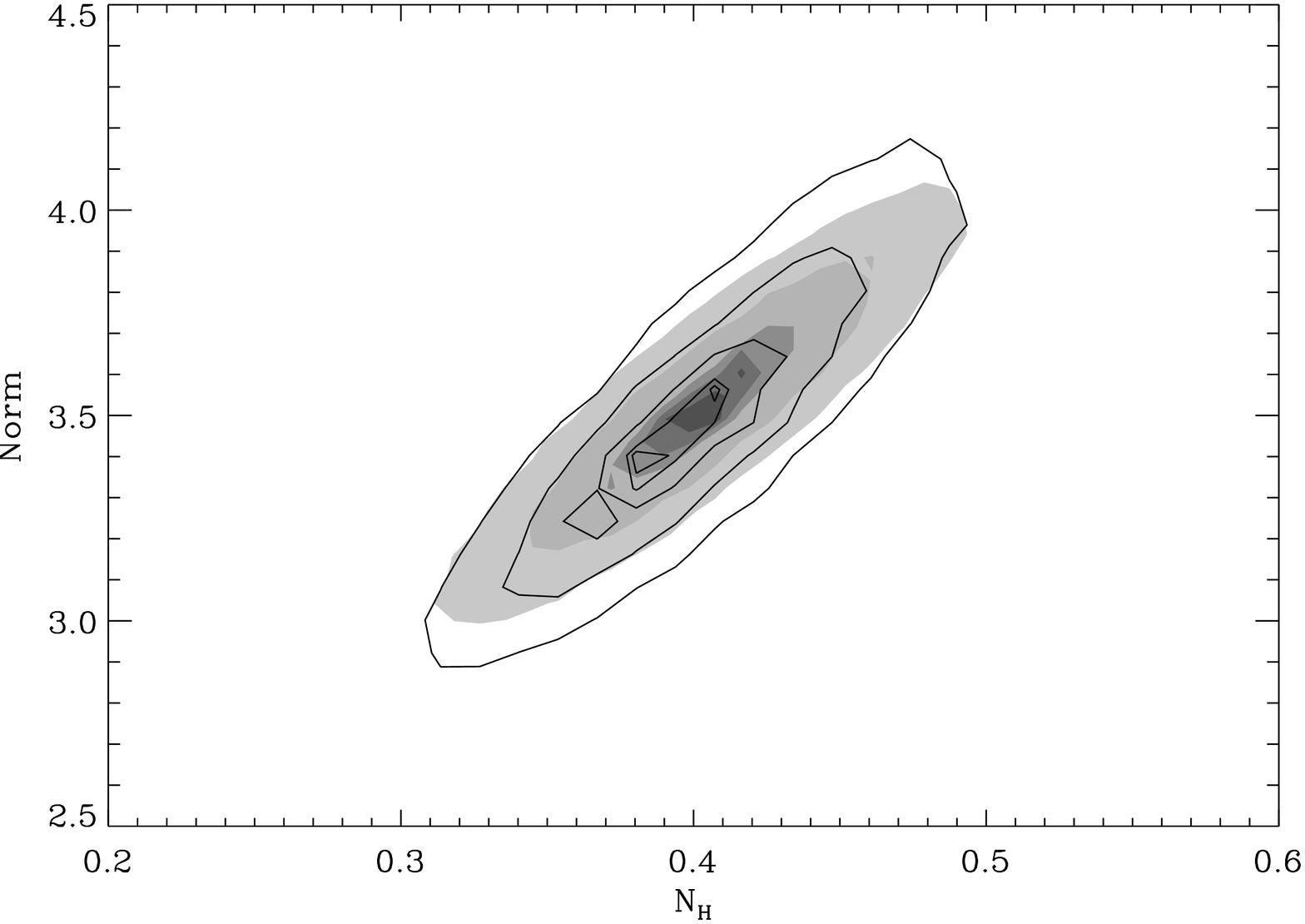}
     \includegraphics[width=5.8cm,angle=90]{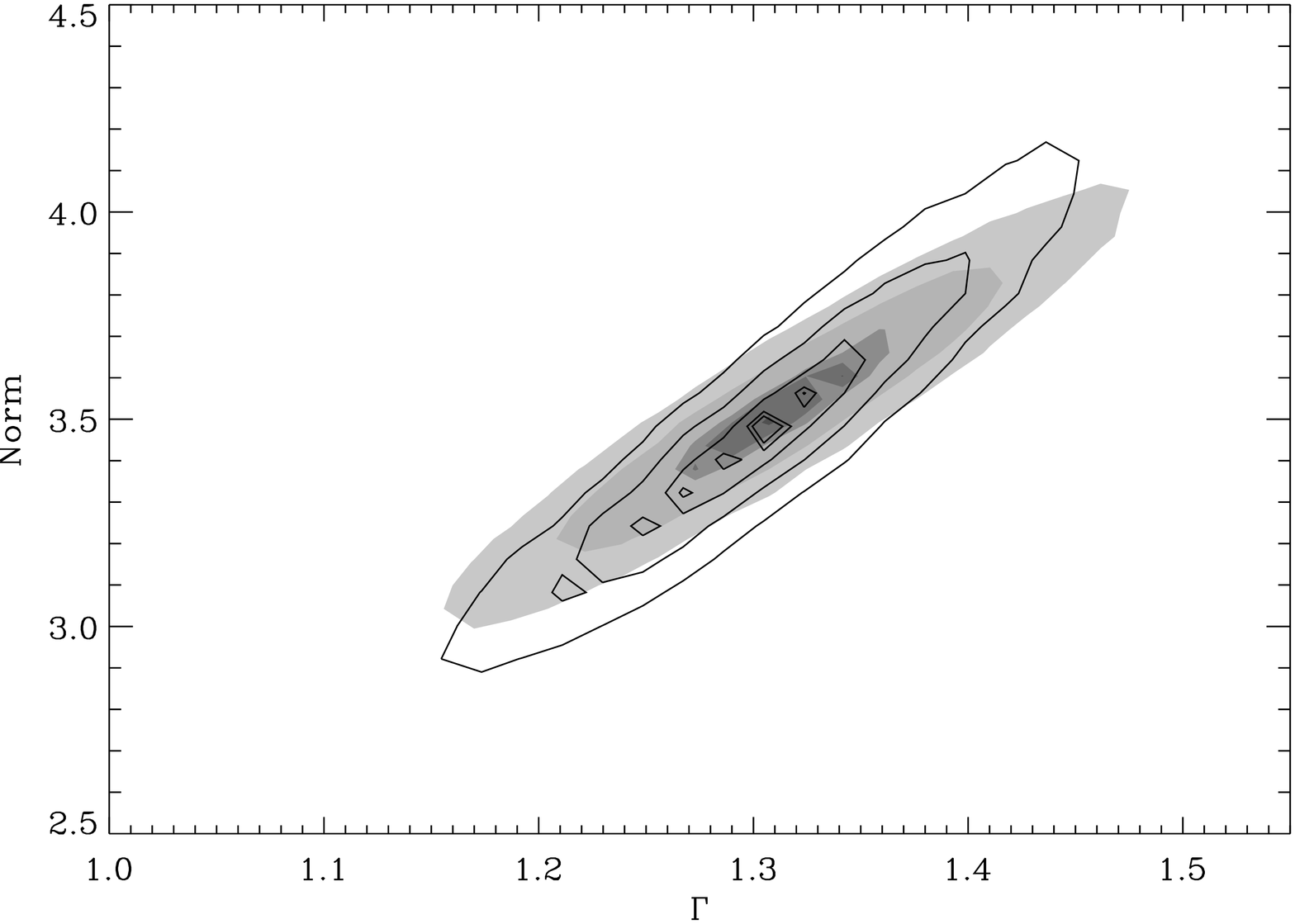}
     \includegraphics[width=5.8cm,angle=90]{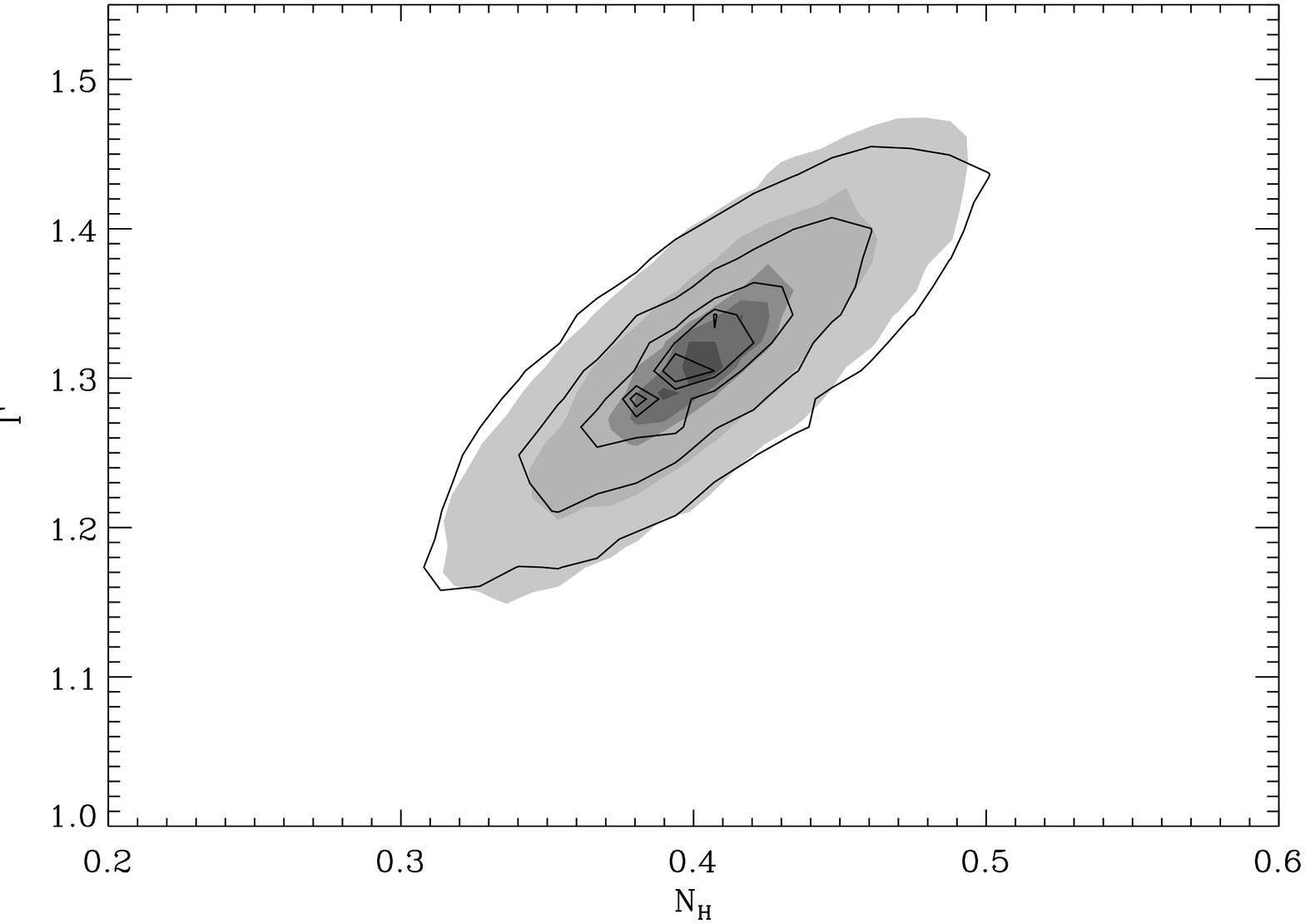}
     \caption{\label{fig:mcmc} 
       Marginal posterior distributions for the continuum parameters
       of an absorbed power law fit to an XRT spectrum of GRB 060124.
       The contours enclose $80$, $50$, $20$, $10$ and $5\%$ 
       of the distribution (and therefore correspond to
       $20$, $50$, $80$, $90$, $95\%$ credible regions for the parameters).
       The filled contours were computed assuming the posterior
       is a Gaussian.
       The hollow contours were computed using MCMC simulations
       from the routine of \citet{vandyk_2001}.
       The two distributions are clearly very similar.
       See section~\ref{sect:approx} for details.
       }
\end{center}
\end{figure}

As a further test of the Gaussian approximation we compared the
posterior calculated using the Laplace approximation with the posterior
calculated using the MCMC algorithm discussed in \citet{vandyk_2001}.
The MCMC method does not use an analytical approximation for the
posterior, and therefore is a more general method, but is
computationally demanding.
Figure~\ref{fig:mcmc} illustrates the two posterior distributions
calculated for the specific case of 
a spectrum from GRB 060124. 
The Gaussian data were computed from $10^5$ random draws from a multidimensional Gaussian with
a covariance matrix evaluated as the minimum $\chi^2$ location using
{\it XSPEC}. 
The non-Gaussian data were generated from $10^5$ draws generated\footnote{
Following \citet{vandyk_2001} we generated five seperate chains,
starting from different, `overdispersed' positions within the parameter space
(all outside the $99\%$ confidence region calculated using $\Delta \chi^2$)
and used the $\hat{R}^{1/2}$ statistic to assess their convergence.
We collected data from the chains only after $\hat{R}^{1/2} < 1.01$.
} by
the MCMC routine of \citet{vandyk_2001}.
It is clear that the two distributions are not identical but are very
similar both in terms of size and shape. In the present
context it is important that the ``credible regions'' occupy similar
volumes of parameter space.

The above analyses demonstrate the Gaussian approximation is
reasonable for the posterior of the simple ``continuum'' model $M_0$, 
which is the denominator of equation~\ref{equation:bfactor}. 
The same will be true of the more complex ``continuum $+$ line'' model
$M_1$ when the line is well detected (see section 11.3 of
\citealt{gregory2005}). When the line is weakly detected the posterior
will be close to the boundary of the parameter space, in which case
the Gaussian approximation will not be so accurate.
Indeed, when the MLE of the line normalization is close to the
boundary the likelihood (and
therefore posterior) enclosed in
the allowed region of parameter space will be smaller than that given
by the Laplace approximation, which assumes the Gaussian function extends 
to infinity in all directions. 
This will also happen, for example, when the best-fitting 
line energy is near the limit of the allowed energy range. In such cases there will
be a tendency to overestimate the Bayes factor (i.e. favor $M_1$).
But when the line is weak there may be multiple peaks in the
likelihood (and posterior) which are not accounted for explicitly in
the Laplace approximation. 
We therefore treat the calculated Bayes factors only as a rough guide
to the presence of a spectral line. 

\subsubsection{Assigning Priors}
\label{section:priors}

Bayes factors are sensitive to the choice of prior density.
As stated above, using the Laplace approximation the 
resulting Bayes factors are sensitive to the prior densities only
at the MLE parameter values, but we must exercise care in
assigning prior density functions in order that these values are reasonable.
Fortunately, the prior densities for all parameters that are common to
$M_{0}$ and $M_{1}$ (such as photon index and normalization) are the
same for $M_{0}$  
and $M_{1}$, and therefore cancel out in the ratio. For the other
parameters we have no cogent information except for their allowed
ranges. In such cases we should use the ``least informative'' prior
densities (see e.g. \citealt{Loredo_1990}; \citealt{Sivia_1996}; \citealt{gregory2005} and
references therein for further discussion).

There are wide ranges of possible line energies and redshifts and so 
the line energy, $E_{line}$ is only constrained
to lie within the  useful XRT bandpass, typically $0.3-10$~keV.
We therefore assigned a uniform prior density $p(E_{\rm line}|M_1) =
1/[E_{\rm max}-E_{\rm min}]$ over this range. 
For most spectral fits the line width $W$
was initially held fixed at a value below the instrumental 
resolution\footnote{$\sigma$ = 59 eV (at 5.895 keV) at launch (A. Beardmore, private 
communication)}, and later allowed as a free parameter.
For those models in which the width of the line was a free parameter,
the width was assigned a uniform prior over the allowed range
(usually $0.0-0.7$~keV): $p(W|M_{1}) = 1/[W_{\rm max}-W_{\rm min}]$.

In order to test the dependence of the results to the
prior densities, two non-informative prior assignments
were made for the line strength
(normalization), $A$, following the discussion in  \citet{gregory2005}.
Firstly, following $\S$4.2 of \citet{Sivia_1996} the line strength
was assigned a uniform prior between zero and some upper limit
$A_{\rm max}$. 
Previous reports of emission lines have estimated the line flux $A$ to be as little as a few percent (\citealt{Reeves_2002}; 
\citealt{Watson_2003}) or as much as $\sim 40-80\%$ (\citealt{Yoshida_1999}; \citealt{Piro_2000}) of the total flux. We conservatively
take $A_{\rm max}$ to be the total flux of the spectrum over the evaluated
bandpass (i.e. our constraint is that the line flux is between 0-100\%
the source flux).
However, there are strong arguments 
(\citealt{Loredo_1990}; \citealt{gelman}; \citealt{gregory2005})
that such as `scale' parameter should be given a
{\it Jeffreys prior}, $p(A|M_1) \sim 1/A$, which corresponds to a constant density in
$\log(A)$. Formally this is an improper prior (cannot be
normalized such that its integral is unity), but one can apply
reasonable upper
and lower bounds in order to form a proper prior density.
Following equation 3.38 of \citet{gregory2005} we used
$p(A|M_1) = 1/A \ln[A_{\rm max}/A_{\rm min}]$. In the present context $A_{\rm
  max}/A_{\rm min} = 800$, since a reasonable lower limit to the
X-ray counts from a line is one count, and a reasonable upper limit is 
$800$, the total number of counts in the spectrum. This yields
$p(A|M_1) = 1/6.68A$ as a normalised Jeffreys prior.
The prior density is therefore higher for weaker lines in the Jeffreys case compared to
the uniform case at values (i.e. over $1.25 \times 10^{-3} A_{\rm max}
\le A < 0.15 A_{\rm max}$), and lower for stronger lines.

The ratio of the prior densities at the modes of the two likelihood functions is then 
simply
\begin{eqnarray}
\frac{p(\hat{\bt}_1|M_{1})}{p(\hat{\bt}_{0}|M_{0})} & = &
p(\hat{E}_{\rm line}|M_{1}) p(\hat{W}|M_{1}) p(\hat{A}|M_{1}) \nonumber \\
 & = & \frac{1}{([E_{\rm max}-E_{\rm min}][W_{\rm max}-W_{\rm min}]A_{P})},
\label{equation:ratio_priors}
\end{eqnarray}
in the ranges $E_{\rm line} \in [E_{\rm min},E_{\rm max}]$, $W \in
[W_{\rm min},W_{\rm max}]$ and zero elsewhere.
Here, $A_{P} = A_{\rm max}$ in the uniform case or $A_{P} =
6.68\hat{A}$ in the Jeffreys case.
We have used both uniform and
Jeffrey's priors in the analysis discussed below (see
section~\ref{section:alt-prior}).

\subsection{Posterior Predictive $p$-values (ppp)} \label{Monte_Carlo_method}

The use of posterior predictive $p$-values (\ppp) was advocated, and
demonstrated by application to GRB spectra, by \citet[see
$\S$4.1 for a description of their method and $\S$5 for its
application to GRB 970508]{Protassov_2002}. Like Bayes factors this
method is grounded in Bayesian probability theory. 

One uses the posterior density, $p(\bt | \mathbf{D})$, for the model parameters
conditional on the data -- which defines our state of knowledge about the
parameters given the data and the available prior information --
to determine the posterior predictive distribution -- which is the distribution of
possible future data predicted based on the observed data.
(`Predictive' because it predicts possible future
datasets and `posterior' because the parameters are drawn from the
posterior density of the parameters.)
The posterior predictive distribution is:
\begin{equation}
p(\mathbf{D}^{\rm sim} | \mathbf{D} ) = 
 \int p(\mathbf{D}^{\rm sim}, \bt | \mathbf{D} ) d\bt = 
\int p(\mathbf{D}^{\rm sim} | \bt) p(\bt | \mathbf{D} ) d\bt
\label{equation:ppp_data}
\end{equation}
where $\mathbf{D}^{\rm sim}$ are the possible future datasets (simulations).
In practice the posterior density is used to generate a set of random parameter values
$\bt_i^{\rm sim}$ ($i=1,2,\ldots$) and each of these is used to  
simulate a random dataset $\mathbf{D}_i^{\rm sim}$. The set of simulated data 
from all the possible random parameters  defines the posterior predictive
distribution for simulated data. This in turn can be used to define
the posterior predictive distribution for some test statistic
$T(\mathbf{D})$ (which is a function of the data):
\begin{equation}
p[ T(\mathbf{D}^{\rm sim}) | \mathbf{D}] = \int p[T(\mathbf{D}^{\rm sim}) | \bt] p[\bt | \mathbf{D} ] d\bt
\label{equation:ppp_stat}
\end{equation}
(compare with equation~\ref{equation:ppp_data}).
The posterior predictive $p$-value (\ppp) is the fraction of this
distribution for which $T(\mathbf{D}^{\rm sim}) > T(\mathbf{D})$,
i.e. the area of the tail of the distribution with values of the test
statistic more extreme than the value from the observed data.
\begin{equation}
p = \int_{T(\mathbf{D})}^{\infty} p[ T(\mathbf{D}^{\rm sim}) | \mathbf{D}] d\mathbf{D}^{\rm sim}
\label{equation:ppp_theory}
\end{equation}
where the integration is  taken over the posterior predictive distribution of $\mathbf{D}^{\rm sim}$.
As such the \ppp\ value is a Bayesian analogue of the $p$-value of null hypothesis tests familiar from 
classical statistics (e.g. the $F$ or $\chi^2$ tests). See chapter 6 of \citet{gelman} or \citet{gelman_1996} 
for a general discussion of the \ppp\ method, and \citet{Protassov_2002} for application to GRB data.  

Using the posterior predictive distribution from
equation~\ref{equation:ppp_data} one can produce
a large number of random simulated datasets to be used in a
Monte Carlo scheme to calculate the integral of
equation~\ref{equation:ppp_theory} numerically.  
The steps for a Monte Carlo method for computing 
the posterior predictive distribution to calibrate
the test statistic $T$ are as follows:
\begin{enumerate}

\item Compute the value of the test statistic for the observed data, $T(\mathbf{D})$

\item Randomly draw $N$ sets of $M_0$ model parameter values
  $\bt_i$ for $i=1,2,\ldots,N$ according to the
  appropriate posterior distribution $p(\bt | D)$

\item For each of $i=1,2,\ldots,N$ simulate a dataset $\mathbf{D}_i^{\rm sim}$ using the
  randomly drawn parameter values $\bt_i$. This accounts
  for uncertainties in the parameter values.

\item For each of the simulated datasets compute the test statistic
  $T(\mathbf{D}_i^{\rm sim})$. This is the posterior predictive distribution of
  the test statistic given the observed data $\mathbf{D}$.

\item Compute the posterior predictive $p$-value as the fraction of
  simulated datasets that gave a test statistic more extreme than that
  for the observed data:
\begin{equation}
  p = \frac{1}{N} \sum_{i=1}^{N} \Theta[ T(\mathbf{D}_i^{\rm sim}) - T(\mathbf{D}) ]
\label{equation:ppp}
\end{equation}
where $\Theta$ is the Heaviside step function which simply counts
instances where $T(\mathbf{D}_i^{\rm sim}) > T(\mathbf{D})$.

\end{enumerate}
The number of simulations, $N$, must be large to ensure a good
approximation to the integral of equation~\ref{equation:ppp_theory}
(which is a multiple integral, being itself the integral of the function
computed by equation~\ref{equation:ppp_stat}).
See \citet{Protassov_2002} for more detailed discussion.

\subsubsection{Application to GRB X-ray spectra}
\label{section:ppp-grb}

As discussed above we may approximate the
posterior density for the
parameters, using a multidimensional Gaussian centered on the MLE
values and with a shape defined by the covariance 
matrix evaluated at the peak ($\bs^2$). 
The randomised parameter values needed for step $2$ above may then be
generated with the Cholesky method.

For the purposes of the present paper we use as the test statistic the change in the $\chi^2$
fit statistic\footnote{ \label{footnote:dchi} The $\Delta \chi^2$ statistic is familiar to most X-ray astronomers
and was used in the Bayes factors method above. Here we note that it is equivalent to the likelihood
ratio test (LRT) statistic, since using equation~\ref{equation:logl} we have $\Delta \chi^2 =
\chi_{(0)}^2 - \chi_{(1)}^2 = -2 \ln \lambda$, where $\lambda = p(\mathbf{D} | \hat{\bt}_0, M_0) /
p(\mathbf{D} | \hat{\bt}_1, M_1)$ is the ratio of the likelihood maxima of the two models. Under
the assumptions for which the LRT is valid this should be distributed as $\chi^2$ with degrees of
freedom equal to the number of additional free parameters in model
$M_1$ compared to $M_0$. 
The reason for chosing the LRT  over related statistics,
such as the $F$-test, is that LRT is more powerful.
See \citet{Freeman_1999} and \citet{Protassov_2002} and references therein
for details.} 
between the two models, $M_0$ and $M_1$. This is equivalent to the
formulation discussed in \citet{Protassov_2002}.
The observed data
were fitted with the model $M_0$ and the covariance matrix
evaluated at the best-fit point used to construct the
multivariate Gaussian distribution from which parameter values were
randomly drawn\footnote{In practice this was performed using the 
\texttt{tclout simpars} command in {\it XSPEC\/}}. For each set of model $M_0$
parameter values a spectrum was simulated with the appropriate
response matrix and exposure time, with counts in each channel drawn from a Poisson distribution, and
binned in the same manner as the observed data.

In order to calculate the test statistic for each simulation,
$T(\mathbf{D}_i^{\rm sim})$ it was necessary to fit each simulated
dataset with the two competing models $M_0$ and $M_1$, for each one find the
best-fitting parameters, and compute
$\Delta\chi^{2}_i$. 
This necessarily involves a computationally expensive
multi-dimensional parameter
estimation for each of the $N$ simulations.
We use as standard $N=10^4$ simulations which yields a
$p$-value accurate to four decimal places at very highest and lowest $p$-values
(there is an uncertainty on the \ppp\ value from the finite number of
simulations which is roughly $\sqrt{p(1-p)/N}$ from the binomial
distribution). 
This is acceptable for determining $p$-values as low as $p \sim
10^{-4}$, i.e. $99.99\%$ `significance'. 

As a further test of the validity of the Gaussian assumption for the
posterior (see also sections~\ref{section:applic}-\ref{sect:approx})
we have compared  results with and without this assumption.
In particular, we calculated the \ppp-value for a spectrum of GRB
060124 using Gaussian parameter values and also using values generated
by the MCMC method discussed by \citet{vandyk_2001}. The two results
were reasonably close ($p = 0.050 \pm 0.007$ from the Gaussian
simulations and $p = 0.073 \pm 0.008$ from the MCMC, based on $10^3$
simulations). This confirms the point made in section~\ref{sect:approx}, that 
the Gaussian assumption is reasonable for these data.

\subsubsection{Automated fitting of GRB spectra}
\label{section:fitting}

Given the number of simulated datasets one must resort to an automated
model fitting procedure. This has itself been the cause of some
debate, with some authors (e.g $\S$5 of \citet{Rutledge_2003})
claiming that automatic routines do not robustly find the best-fitting
parameter values (minimum $\chi^2$).  The algorithm used by {\it
XSPEC} for $\chi^2$ minimization is the Levenberg-Marquardt algorithm,
which is efficient and very effective when the $\chi^2$ space is
well-behaved (e.g. with only one local minimum). However, as this is a
`local' routine there is no guarantee of finding the `global' minimum
in $\chi^2$, and it is possible that the results are biased by the
presence of other local minima.  For the present paper we have
employed several additions to the standard Levenberg-Marquardt
minimization algorithm in order to mitigate these problems.

Once a local minimum in $\chi^2$ is found the
surrounding region of parameter space is explored for signs of other
minima.  Each parameter in turn has its value increased and decreased
until the $\chi^2$ is increased by at least $\Delta \chi^2 = 2.7$,
while simultaneously allowing the other parameters to vary in order to
minimize $\Delta \chi^2$. If any non-monotonicity in $\chi^{2}$ is
detected during this search the volume of parameter space explored is
increased by increasing the value of $\Delta \chi^2$. If during the course
of this search $\Delta \chi^2$ becomes negative (meaning there is a
lower minimum nearby) the Levenberg-Marquardt algorithm is re-started
from the position of this new minimum.  The entire process is repeated
by perturbing each parameter in this way until no further improvement
can be made by the adjustment of any of them.  

The absorbed power law model ($M_0$) has only three parameters (photon
index $\Gamma$, normalization and absorption column density), and in
all cases finding the $\chi^2$ minimum was straightforward using the
above procedure. The
alternative model $M_1$, which includes the emission line, required
more care because the presence of a line with unknown energy may cause local
$\chi^2$ minima at different energies within the wide bandpass. 
An initial `best guess' line energy was computed for each spectrum
in the following way. An absorbed power law plus emission line model
was constructed using the best-fitting parameters of model $M_0$ and adding an unresolved emission
line fixed at some trial energy $E_{i}$ and varying the other parameters (including
the line normalization) to find the minimum $\chi^2$. One hundred
values of the trial energy $E_{i}$ were used, evenly spread over the
entire useful bandpass, and the value that recorded the lowest
$\chi^2$ was selected as the `best guess' for the line energy. The
enhanced Levenberg-Marquardt algorithm described above was then used
to find the global $\chi^2$ minimum starting from this position.

Simulation tests and
comparison with interactive fitting demonstrated the automatic
procedure described above was an
efficient and very robust procedure for finding the global minimum.

\subsection{Rutledge and Sako Smoothing (RS)} \label{subsection:R_n_S_analysis_method}

\citet{Rutledge_2003} proposed an alternative method for line
detection using a `matched filter' to smooth the 
observed count spectrum with the aim of removing low significance noise and
emphasizing any spectral features.
The distribution of peak fluxes in the smoothed spectrum is then
compared to the result of Monte Carlo simulations to calibrate their
significance ($p$-value).

The counts per PHA channel are extracted from the observed X-ray
spectrum and then smoothed using an energy-dependent kernel 
(a Gaussian having a FWHM equal to the spectral resolution of the
detector; see equation 2 of \citet{Rutledge_2003}) to produce the smoothed spectrum $C(E)$. 
The distribution of $C(E)$ is then calibrated using Monte Carlo simulations of spectra generated using the 
method discussed in section~\ref{Monte_Carlo_method} that employs posterior predictive data sets.
(We note that \citealt{Rutledge_2003} and \citealt{Sako_2005} did not
randomize the parameter values but used fixed MLE values to generate
all their simulations. This is equivalent to assuming the posterior to be
a delta function located at the best fit point, which is clearly a
bad approximation in many cases.)
Each simulation is in turn smoothed using the same energy 
kernel to produce $C(E)_{sim,i}$. The $C(E)_{sim}$ values are then
sorted in descending order for each PHA channel separately. Thus the
99$^{th}$  
percentile limit of the $C(E)_{sim,global}$ is then found by
extracting the 100$^{th}$ highest value of $C(E)_{sim}$ in each PHA
channel. 

The smoothed observed spectrum, $C(E)$, is then plotted alongside the
$n^{th}$ percentile limits, which we have chosen for this analysis  
to be 90.00, 99.00, 99.90 and 99.99 $\%$. Wherever $C(E)$ exceeds a
given limit then we have detected a `feature' at that confidence
limit. 
Thus a line would show up as a narrow excess whilst other thermal
emission components will show up as broad excess, both of which are  
easily distinguishable.

\subsection{Comparison of the methods}
\label{section:compare}

The three methods discussed above have different theoretical
motivations, underlying assumptions and require different amounts of
computing power. The Bayes factor method is based on a simple
application of Bayes theorem combined with the Laplace approximation
and assumes uniform priors on the model parameters (or Jeffreys prior
for the line normalization). As discussed
above, this may not be the optimal assignment. However, despite its 
possible drawbacks, the simple priors and Laplace approximation make
the calculation extremely simple,
requiring only the evaluation of equations~\ref{equation:odds_2} and
\ref{equation:ratio_priors} which require the
values of $\chi^2$ and the covariance
matrices for the best-fitting line and line-free models, and
details of the free parameters and their allowed ranges. As such, it
is useful as a `quick and easy' test. The dependence of choice of
priors may be assessed by comparing the results computed using the
uniform and Jeffreys prior.

By contrast, the RS and \ppp\ methods require a large number of random
datasets to be simulated and analyzed, and are therefore considerably
more costly in terms of computing time. There is no compelling
theoretical reason for applying a matched filter, as in the RS method,
although it should be noted that the method, as implemented above, is
calibrated using the appropriate posterior predictive
distribution. The advantage of the RS method is that no model fitting
is required, which is often a time-consuming process and can lead to
biased results if not handled properly (section~\ref{section:fitting}).

The \ppp\ method is grounded in the theory of Bayesian model checking
(\citealt{gelman}; \citealt{Protassov_2002}) but requires
time-consuming fits to be  performed on each simulated spectrum, and
is therefore the most computationally demanding method by a clear
margin. However, it is arguably the most rigorous in the sense that it
is less sensitive  to the choice of priors than are Bayes factors
\citep{gelman_1996, Protassov_2002}, and does not apply an ad hoc
smoothing, as in the RS method, that may actually act to suppress real
spectral features in  some cases.

The simulations used for both RS and \ppp\ methods were generated
assuming a Gaussian posterior for the three parameters of $M_0$,
which, as discussed in section~\ref{sect:approx} is a good
approximation. Again, this approximation was made to increase
computational efficiency, since Gaussian deviates are trivial to generate
with the Cholesky method. 
In situations where the Gaussian approximation is not valid and/or 
the number of spectra is small enough that considerably more computing
time may be spent on each, the \ppp\ method or Bayes factors may
be computed using results from MCMC simulations \citep{vandyk_2001,
  Protassov_2002} which allows for a more accurate evaluation of the
posterior density. 

\subsubsection{Alternative approximate methods}

The statistics literature contains many methods developed for the
purpose of model selection. In the introduction  we listed four
methods that have previously been applied to the problem of line
detection in X-ray data from GRBs.  One method that has not, to our
knowledge, been applied specifically to GRB line detection is the
Bayesian Information  Criterion (BIC; \citet{schwartz_1978}). This
aims to approximate the logarithm of the integrated posterior
probability  for a model with $k$ parameters given data with a sample
size $N$. The BIC takes the form of the logarithm of the  likelihood
with a penalty term:

\begin{equation}
\label{eqn:bic}
BIC = - \ln [ p(\mathbf{D}|\bt, M)] + (k/2) \ln N
\end{equation}

The model with the smallest BIC is favored. The difference between the
BIC values for two competing models (often called  the \emph{Schwartz
criterion}) is therefore $S = - \ln \lambda + (\Delta k/2) \ln N$ (see
footnote~\ref{footnote:dchi}), and is a rough approximation to the
logarithm of the Bayes factor (section 4.1.3 of \citealt{Kass}).

In the high count (large sample size) limit (see
equation~\ref{equation:logl}) the Schwartz criterion becomes  $2S = -
\Delta \chi^2 + \Delta k \ln N$. Whether or not the BIC for model
$M_1$ is smaller than that for $M_0$ is then equivalent  to the
criterion $\Delta \chi^2 > - \Delta k \ln N $. In the present case the
data are selected with fixed $N$, and $\Delta k = -2$  for the
addition of a fixed width line, such that the BIC is equivalent to
applying the same $\Delta \chi^2$ criterion to each spectrum,
mechanically the same as the LRT, although with a different (generally
higher) threshold value. Therefore, in the  present context the
application of the BIC would be equivalent to a slightly more
conservative application the LRT  (see
footnote~\ref{footnote:lrt}). However, as noted in
\citet{Protassov_2002} and elsewhere, the BIC is often a poor
approximation  to the integrated posterior probability, and as
discussed by \cite{Kass} is generally a worse approximation than the
Laplace  approximation employed to calculate Bayes Factors in
section~\ref{section:applic}.

\section{Results from an Iron line emitting source} 
\label{section:PKS}

%In addition to testing fake spectra with our analysis methods we also applied them to an existing {\it Swift\/} PC mode calibration dataset 
%(combining all available data from 10/05/2005 to 02/09/2005) of PKS 0745-19 (\citet{DeGrandi_1999} and \citet{Chen_2003}). This is a galaxy 
%cluster with a thermal spectrum and a known line at 6.07 keV in {\it Swift's\/} observations, 
%which is a redshifted 6.7 keV iron line ($z = 0.1$). Even though the underlying spectrum is thermal, with multiple temperature components, it can be modelled 
%by an absorbed powerlaw continuum where the powerlaw index, $\Gamma$, is $2.34^{+0.03}_{-0.03}$ ($\chi^{2}/\nu = 635/511$). An additional mekal component, 
%with kT = $0.19^{+0.03}_{-0.01}$, slightly improved the fit with $\chi^{2}/\nu = 610/509$. All spectral parameter errors are quoted at 90$\%$ confidence. 

As a first demonstration of the above methods we applied them to a non-GRB {\it Swift\/} 
dataset. Ideally we would prefer to examine a source with a GRB-like spectrum, with a 
similar count rate, but containing a clearly identified emission line feature. However, 
it is difficult to find a source that meets all of these criteria. We chose the PC mode 
calibration dataset (combining all available data from 10/05/2005 to 02/09/2005) of PKS 0745-19 
(\citet{DeGrandi_1999} and \citet{Chen_2003}). This test has some limitations as PKS 0745-19 is 
fainter than the GRBs analyzed in this paper and it is observed in a different mode. 

PKS 0745-19 is a galaxy cluster with a thermal spectrum and a known line at 6.07 keV in 
{\it Swift's\/} observations, which is a redshifted 6.7 keV iron line ($z = 0.1$). Even 
though the underlying spectrum is thermal, with multiple temperature components, it can 
be modeled by an absorbed power law continuum where the power law index, $\Gamma$, is 
$2.34^{+0.03}_{-0.03}$ ($\chi^{2}/\nu = 635/511$). An additional \texttt{mekal} 
(\citet{Mewe_1985} and \citet{Arnaud_1996}) component, with kT = $0.19^{+0.03}_{-0.01}$, 
slightly improved the fit with $\chi^{2}/\nu = 610/509$. All spectral parameter errors are 
quoted at 90$\%$ confidence. 

\begin{figure}
\begin{center}
     \includegraphics[width=8.5cm,angle=270]{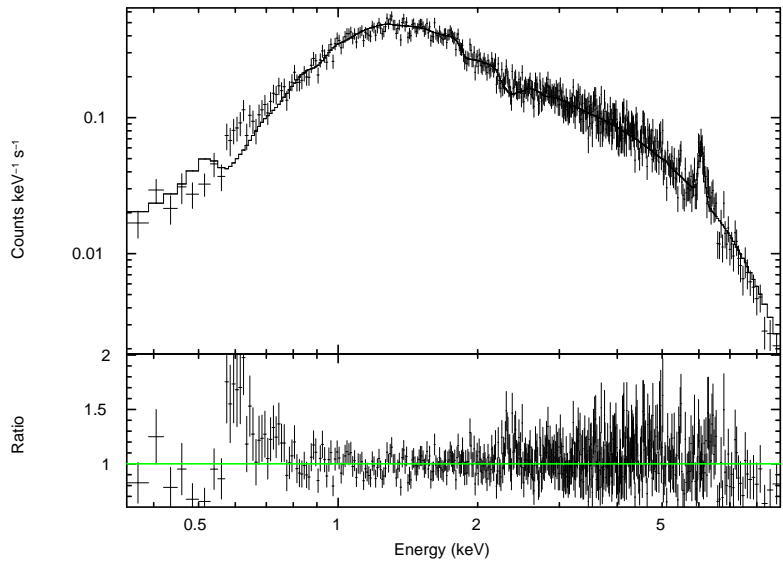}
     \caption{\label{figure:PKS0745_spec} Spectral fit to PKS 0745-19 with an absorbed power law plus a narrow Gaussian emission line model. The redshifted Iron line at 6.07 keV is clearly visible. Note also the residuals at 0.6 keV and 2.3 keV, which are thought to be due to the oxygen and gold edges respectively.}
\end{center}
\end{figure}

\begin{figure}
\begin{center}
     \includegraphics[width=8cm,angle=270]{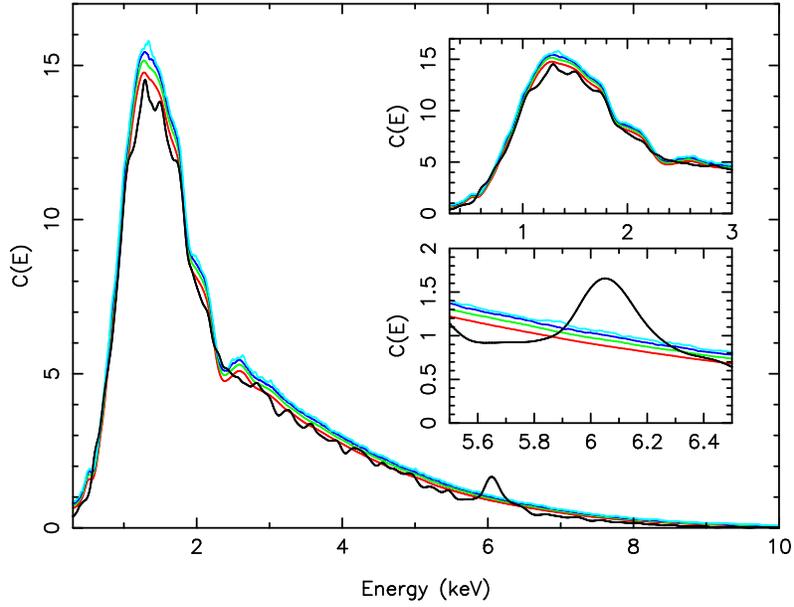}
     \caption{\label{figure:PKS0745_smoothed} PKS 0745-19: RS method. Confidence contours mark the significance of the spectral features. 
Red $= 90.0\%$, green $= 99.0\%$, dark blue $= 99.90\%$ and light blue $= 99.99\%$. Insets focus on energy ranges of interest.}
\end{center}
\end{figure}

Adding a Gaussian component to an absorbed power law fit naturally produced a significantly 
improved fit to the data ($\chi^{2}/\nu = 539/508$) with $\Gamma = 2.36^{+0.04}_{-0.03}$ and 
a line at $6.07^{+0.02}_{-0.02}$ keV (width = $0.06^{+0.02}_{-0.03}$ keV). The spectral fit to 
this model can be seen in fig~\ref{figure:PKS0745_spec}. This is supported by the Bayes factor 
of $7\times10^{14}$ for a single line being present. RS analysis of the spectrum, 
fig.~\ref{figure:PKS0745_smoothed}, also clearly showed the presence of a Gaussian feature at 
$\sim 6.07$ keV with a significance far in excess of the $99.99\%$ confidence limit. The \ppp\ 
analysis placed a significance of $>99.99\%$ on this feature.

An interesting point to note is that there are shallow `excesses' at $\sim 0.6$ keV and $\sim2.3$ 
keV in the RS plot (fig~\ref{figure:PKS0745_smoothed}), which are clearly not line features. Coherent, 
low level, positive excesses are also seen in the spectral fit at these energies (fig.~\ref{figure:PKS0745_spec}). 
Either the power law component is not modeling the data adequately at these points, the energy scale 
for this spectrum has an offset or the calibration files are less accurate around these two energies. 
Applying the \texttt{gain fit} function in {\it XSPEC\/} improves the model fits significantly by adding 
an offset\footnote{http://swift.gsfc.nasa.gov/docs/heasarc/caldb/swift/docs/xrt/xrt$\_$bias.pdf} 
of -0.07 keV (no change to the slope). The absorbed power law model improves from $\chi^{2}/\nu = 635/511$ 
to $606/509$ and the \texttt{mekal} component model improves from $\chi^{2}/\nu = 610/509$ to $585/507$. 
As a result the two shallow `excesses' at $\sim 0.6$ keV and $\sim2.3$ keV become far less prominent.

The feature at $\sim 0.6$ keV could be attributed to the detector oxygen absorption edge at 0.54 keV. 
Applying the -0.07 keV offset brings the $\sim 0.6$ keV `line' in conjunction with this edge, thus 
reducing its significance below the point at which we would consider it to be a real detection. We note 
that the $\sim 2.3$ keV feature is coincident in energy with the gold edge due to the XRT mirrors. We have 
confirmed that this feature is not due to any bad pixel or hot column 
issues\footnote{http://swift.gsfc.nasa.gov/docs/heasarc/caldb/swift/docs/xrt/SWIFT-XRT-CALDB-01$\_$v5.pdf}.

\section{Testing the three methods and determining detection limits.} 
\label{section:Detection_limits}

In this section we discuss the sensitivity limits of the three
methods, i.e. the weakest lines that can be reliably detected with
each of the three methods, for observations of the type discussed in
section~\ref{section:Data_reduction}, of a `typical' {\it Swift} era
burst. This is done by simulating XRT data with a continuum
spectral model typical of the GRBs observed with {\it Swift}, but
including an emission line, and then applying the three methods described
above for line detection. 

In order to generate the simulated data we use a fiducial spectral
model comprising a power law 
with photon index $\Gamma=2.0$, normalization (at 1 keV) of $N=0.9$
photons keV$^{-1}$ s$^{-1}$ and an absorption column density of $N_{\rm
  H}=1.8\times10^{21}$ cm$^{-2}$ (see table 2
of \citealt{Campana_2006_AA}). These parameters are typical of the
X-ray spectra of {\it Swift\/} era bursts\footnote{We 
have assumed a redshift $z=0$ for the fiducial burst spectrum.
The average of the measured redshifts for {\it Swift\/} GRBs is 
higher than this (see http://www.astro.ku.dk/$\sim$pallja/GRBsample.html
for the updated values).
However, it should be noted that increasing $z$ causes the effects of
absorption by the host galaxy
absorption (which tends to dominate the total absorption column) to
shift out of the observed bandpass, meaning there is relatively more
flux at lower energies ($<1$ keV). The calculated detection limits
should be representative of {\it Swift\/} era bursts although
perhaps conservative at lower energies.}.
In order to measure the sensitivity of the three detection
methods to lines in XRT data, spectral data were simulated using the above model plus one
Gaussian emission line, and subjected to each of the three procedures. 
A range of values for line energy, normalization and intrinsic width were
used in order to calibrate the dependence of the methods to the line
parameters\footnote{
The ranges of values used for the line simulations are as
follows:
Normalizations of $1\times10^{-7} \to 100$ photons cm$^{-2}$ s$^{-1}$
taken in logarithmically increasing steps,
line energies of 0.4, 0.6, 0.8, 1.0, 2.0, 3.0, 4.0, 5.0, 7.0 and 9.0
keV, and intrinsic widths of $0.0$~keV (i.e. unresolved), $0.2$~keV (broad
line) and $0.7$~keV (broad continuum excess).
}.

The \ppp\ and RS method result in $p$-values with the conventional frequentist interpretation. 
If we set the detection threshold at $\alpha$, and identify a detection as $p \le \alpha$ then 
the rate of type I errors (i.e. false positive detections) will be $\alpha$. For the purpose of  
sensitivity analysis we used $\alpha = 0.01$, equivalent to a ``99\% significance" criterion.
In contrast to these, the Bayes factor is the ratio of the marginal likelihoods of models $M_1$ 
and $M_0$; in the case of uniform priors for the two models this is the ratio of posterior 
probabilities $B_{10} = p(M_1|\mathbf{D})/p(M_0|\mathbf{D})$ where the probabilities are 
interpreted directly as probabilities for models $M_0$ and $M_1$, respectively.

For the purpose of numerical comparison with the $p$-values, the Bayes factors were converted
into probabilities (assuming $p(M_1|\mathbf{D}) + p(M_0|\mathbf{D}) = 1$; see equation 3.19 of
\citet{gregory2005}), and $p(M_0|\mathbf{D}) < 0.01$ was taken as the criterion for
detection. This is equivalent to $p(M_1|\mathbf{D}) > 0.99$, and approximately equivalent to
a Bayes factor $B_{10} > 100$, which is conventionally taken as strong evidence in favor of
$M_1$ over $M_0$ \citep{Kass}. However, we stress that the interpretation of
$p$-values and Bayes factors are fundamentally different. A $p$-value is the tail area of the
probability density function of the test statistic, assuming a null hypothesis ($M_0$) is
true, and is used to decide whether or not to reject the hypothesis. As such, a $p$-value is
not the probability for the model $M_0$, instead it corresponds to the frequency of more
extreme test statistics (e.g. $\Delta \chi^2$) given a large number of repeat experiments
(assuming the null hypothesis). By contrast, $p(M_0|\mathbf{D})$ is the posterior probability
for model $M_0$ based on data $\mathbf{D}$ and the priors (in the present case we used an
approximation thereof), as $p(M_1|\mathbf{D})$ is for $M_1$, and Bayes factors are used to
select between two models based on the ratio of these two. This fundemental difference in 
the interpretation of Bayes factors means there is no expectation that $\alpha$ is the 
frequency of type I errors from a large number of repeat observations when using a 
$p(M_0|\mathbf{D}) < \alpha$ criterion.

In $\S$~\ref{sect:approx} we confirmed that using the Laplace approximation assumption in the calculation
of the Bayes factor was valid for the fiducial absorbed power law spectral model. The same was also 
found to be true of the spectra with simulated Gaussian lines at, and above, the detection limit 
detailed above.  

For each value of the line normalization we calculated the Bayes
factors for $50$ independent simulations and calculated the $p(M_0|\mathbf{D})$
values for each. We then averaged the $p(M_0|\mathbf{D})$ values at each
normalization and linearly interpolated between points at adjacent
normalization values to map $p_0$ as a function of normalization.
The limiting sensitivity was taken to be the normalization at
which the mean $p(M_0|\mathbf{D})$ value falls below $0.01$.

\begin{figure}
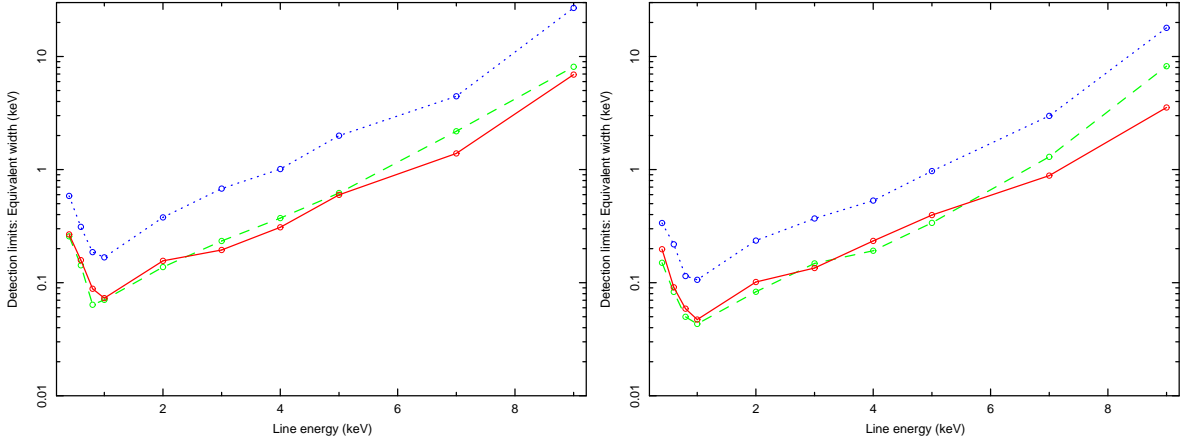

\begin{center}
     \includegraphics[width=5.8cm,angle=270]{f4a.ps}
     \includegraphics[width=5.8cm,angle=270]{f4b.ps}
     \caption{\label{figure:width0.0} {\it Narrow Gaussian line (width
     $<$ instrumental resolution) for a spectrum containing 800
     counts (left) and 1600 counts (right).\/} Comparison of the
     detection limits, in equivalent width (keV), of the three methods
     over the energy band pass of {\it Swift\/}. The data are as
     follows; dotted blue - Bayes factor analysis, solid red -
     RS method, and dashed green - posterior
     predictive $p$-value analysis. }
\end{center}
\end{figure}

\begin{figure}
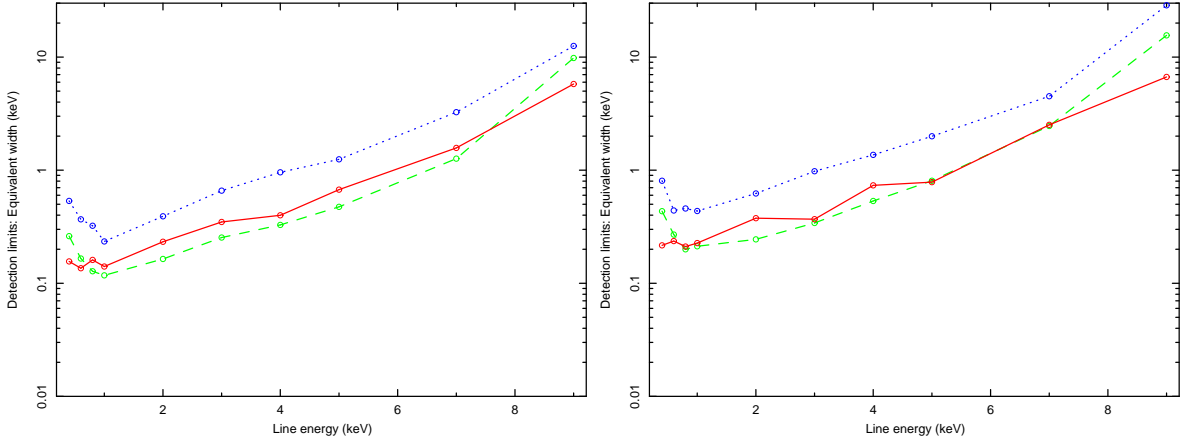

\begin{center}
     \includegraphics[width=5.8cm,angle=270]{f5a.ps}
     \includegraphics[width=5.8cm,angle=270]{f5b.ps}
     \caption{\label{figure:width0.2} {\it Broad Gaussian line (width
     = 0.2 keV) for a spectrum containing 800 counts (left) and 1600
     counts (right).\/} Comparison of the detection limits, in
     equivalent width (keV), of the three methods over the energy band
     pass of {\it Swift\/}. The data are as follows; dotted blue -
     Bayesian analysis, solid red - RS method, and
     dashed green - \ppp\ . } 
\end{center}
\end{figure}

\begin{figure}
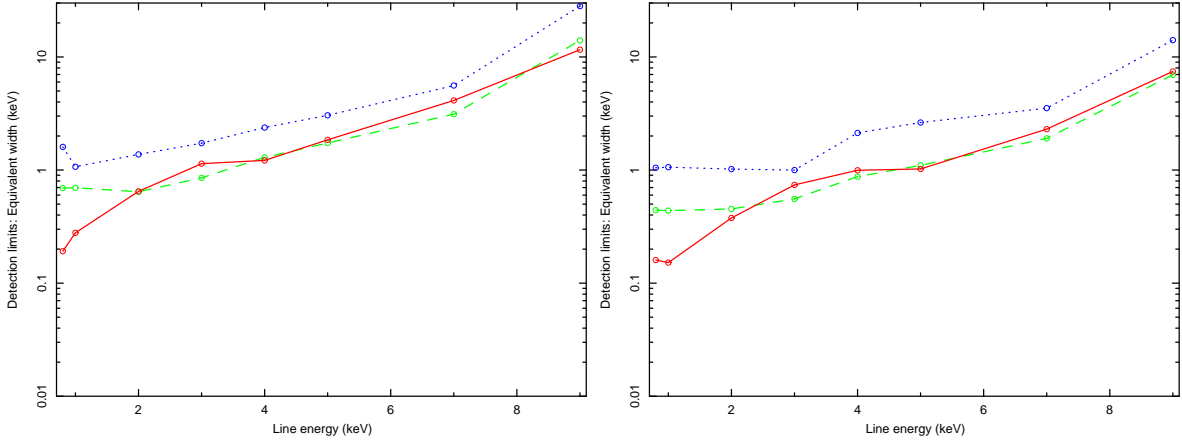

\begin{center}
     \includegraphics[width=5.8cm,angle=270]{f6a.ps}
     \includegraphics[width=5.8cm,angle=270]{f6b.ps}
     \caption{\label{figure:width0.7} {\it Broad excess (width = 0.7
     keV) for a spectrum containing 800 counts (left) and 1600 counts
     (right).\/} Comparison of the detection limits, in equivalent
     width (keV), of the three methods over the energy band pass of
     {\it Swift\/}. The data are as follows; dotted blue - Bayes
     factor analysis, solid red - RS method, and dashed green - \ppp\ . Values below 0.7 keV have been excluded due to the width of the features being analyzed.}
\end{center}
\end{figure}

Figure~\ref{figure:width0.0} shows the detection limits for an
intrinsically narrow line ($W = 0$) at different energies for spectra with $\sim 800$
and $\sim 1600$ counts (left and right panels, respectively). 
The limiting sensitivities are shown in units
of equivalent width (keV), which is easier to interpret physically,
than the absolute normalization, by comparing the normalization to
the underlying continuum model.
Figures~\ref{figure:width0.2} and \ref{figure:width0.7} show the detection limits 
for different line widths ($W=0.2$ and $0.7$~keV, respectively).
The Bayes factor points in these figures have been calculated using the uniform prior, rather
than the Jeffreys prior. See $\S$~\ref{section:alt-prior} for further discussion on the effect of using the two different 
priors in the calcualtion of the Bayes factors for the observed data sets.

The \ppp\ and RS methods require a large number of spectral
simulations in order to calibrate their distribution and estimate the $p$-value.
The computational demands of this\footnote{To give a specific example, for the
simulation and fitting methods described in
sections~\ref{section:ppp-grb} and \ref{section:fitting} a set of
$N=10^4$ simulations takes $\sim 1$ day on a top-range PC.}
are such that it was not feasible to produce a sufficiently large set of simulations 
to carry out the methods on each and every line spectrum (e.g. which includes several
spectra at each trial value of line energy,
width and normalization, for both $\sim 800$ and $\sim 1600$ count spectra).
We therefore constructed two libraries of $10^4$ simulations, one for
$\sim 800$ and one for $\sim 1600$ count spectra, that could be used
for each test. These were constructed by simulating an appropriate
spectrum based on the fiducial model, and using this to generate
the posterior predictive distribution from which to draw $10^4$
simulations following the recipe discussed in
section~\ref{section:applic}. 
These libraries were then used to calibrate the distribution of the $\Delta \chi^2$ statistic 
for the \ppp\ method and thus to calculate the value of $\Delta \chi^{2}$ that corresponds to 
a $p$-value of $0.01$. Similarly these libraries were used to compute the 99.00\% significance 
contour from the fiducial model for the RS method. 
We point out here that these simulation libraries were used only for the purposes 
of comparing the different algorithms. For the analysis of real observations (discussed 
below), each observation was assessed using independently generated simulations matching 
the particular observational parameters. 

For the \ppp\ method each of the spectra containing a line was fitted with an absorbed power 
law with and without an additional Gaussian component, and the change in $\chi^{2}$ noted.
The $\Delta \chi^2$ values were averaged at each normalization, and these points linearly 
interpolated, to map the $\Delta \chi^2$ as a function of normalization. As with the Bayes 
factor, the limiting sensitivity was taken to be the normalization at which the mean $p$-value 
falls below $0.01$, calculated using the appropriated value of $\Delta \chi^{2}$ value from each
simulation library. The limiting sensitivity as a function of energy is shown as green
dotted curves in Figures~\ref{figure:width0.0}, \ref{figure:width0.2} and \ref{figure:width0.7} 
for different configurations of line parameters.

For the RS method each line spectrum was smoothed individually. The $C(E)_{sim}$ values over an 
energy channel range equal to the central energy, $E_{line}$, $\pm$ line width were extracted. 
These values were compared to the 99.0\% confidence limit over the same energy channel range 
found from the appropriate simulation library. The number of channels within this range where 
$C(E)_{sim} > C(E)_{99.0}$ was recorded for each simulation. The detection limit was taken to be
the lowest line normalization where $N(C(E)_{sim} > C(E)_{99.0}) = 0$.

Analysis of the (line free) library simulations showed the Bayes factors produced $< 1\%$ false 
positives when  $B_{10} \ge 100$ was used as a detection criterion. This shows the method is, if 
anything, slightly conservative as expected given the conservative $A_{\rm max}$ assumption. 
Conversely, the false negative detection rate is negligible above the detection limit. 

As expected the detection limits are higher for the spectrum with a lower 
number of counts, by a factor of $\sim$1.5. 
For the fiducial spectral
model used here the optimum energy range for detecting lines 
is $0.4 - 6$ keV, where the line only requires a contribution of a few
$\%$ of the total spectral flux,
In the best cases ($1600$ counts and narrow line) a line with an
equivalent with as small as $\sim 40$~eV may be detected around $1$
keV (observed frame) at $99$\% significance (in a single trial), 
whereas only very strong lines may be detected between $6 - 10$ keV.
Additional simulations were carried 
out with a higher absorption column density ($1.0\times10^{22}$
cm$^{-2}$; the mean values stated in 
\citet{Reichart_2002} assuming that long bursts occur in molecular
clouds). The dependence of line detection with respect to energy for
all three methods were the same at energies $>$1 keV.  
However we note that simulating the spectra with much larger
absorption columns significantly degraded the ability to detect lines
features below 1 keV.  

\section{Results from {\it Swift\/} archival GRB afterglow data} \label{section:Swift}

Our sample covers a subset of 40 GRBs, out of the total of 153 from
GRB 050128 up to GRB 060510B, which were selected for the quality of
their WT mode  
data (see $\S$~\ref{section:Data_reduction}). 
Some bursts only contained sufficient data for a single WT mode
spectrum to be analyzed, whilst the majority contained sufficient data
to be time-sliced  
into multiple spectra (see $\S$~\ref{section:Data_reduction}). In
total 332 spectra were analyzed.
We sample a range of energies and time spans even though the complete
redshift distribution of this 
data set is unknown. The subset of this sample with known $z$
indicates that we are typically probing the region between T+0 s to
T+$\sim$500 s (or up to T+few ks  
if the burst is very bright) post trigger and between $\sim1.0$ and
$\sim50$ keV in the rest frame of the burst. Throughout this section
error bars indicate nominal $90$\% confidence limits on one
interesting parameter. 

All the data were fitted using automated procedure described in section~\ref{section:fitting} and the solutions checked by hand.
In practice four models were fitted to each spectrum: 
(1) absorbed power law; 
(2) absorbed power law plus unresolved Gaussian emission line; 
(3) absorbed power law plus variable-width line; 
(4) absorbed power law plus blackbody. 
The results presented below focus on the line models, and we found that the blackbody parameters were in general very poorly
constrained. Ideally, we would like to apply all three methods to all $332$ WT spectra to assess the significance of lines (or other) 
features in the data. But, as discussed above, the RS and especially \ppp\ methods are computationally demanding and so it was
not practical to apply these methods to every spectrum. 

The (approximate) Bayes factor method, being computationally
economical, was applied to every spectrum, while the more
computationally expensive RS and \ppp\ methods were applied only to
subsets of the data. In particular, any spectrum
that showed a Bayes factor $B_{10} \ge 1$ (in favor of a line), or a
$\Delta \chi^2 \ge 4.61$ upon inclusion of a
line\footnote{\label{footnote:lrt}$\Delta \chi^2 = 4.60517$ is the
$90$th  percentile for the $\chi^2$ distribution with $2$ degrees of
freedom \citep{Press}. As such, it corresponds to a $90$\% detection
`significance' ($p < 0.10$) in a classical likelihood ratio test (LRT)
when including two additional parameters (see
footnote~\ref{footnote:dchi}). The LRT should not be used directly
for the purposes of detecting an emission line (for reasons discussed
in \citealt{Protassov_2002}), but in practice the $p$-value
calculated from the analytical test is usually within an order of
magnitude of the value calibrated using the \ppp\ method. It is
therefore extremely unlikely that a dataset producing $\Delta \chi^2 <
4.61$ would yield a solid detection (e.g. $p < 10^{-3}$)  after \ppp\
analysis.}, was considered for more detailed analysis. These were
deliberately chosen to be extremely relaxed selection criteria
(especially so given the large number of independent tests, see
below), so as to avoid removing any plausible line candidates and only
remove those spectra without any hint of a line, and to counteract the
conservative nature of the Bayes factors
(section~\ref{section:Detection_limits}). Indeed, this 
screening effectively reduced by a factor $\sim 4$ the sample of
spectra worth considering in more detail. We re-iterate that no
judgement about the presence/absence of a line in a spectrum was made
purely on the basis of the Bayes  factor method, which, as discussed
above, is an approximation and is sensitive to the choice of priors. Only spectra with a low
Bayes factor ($B_{10} <1$) \textit{and} little improvement in the fit
statistic upon including a line ($\Delta \chi^2 < 4.61$) were not
considered for further analysis. This subsample was  then subjected to
the RS method with a low detection threshold ($p < 0.1$, i.e. a $90$\%
single trial significance, again very weak given the  multiple trial
effect). This further reduced the sample size to a level where the
rigorous but computationally expensive \ppp\ method could be applied.

As stated above, this screening was only necessary to reduce the
sample to a manageable size for \ppp\ analysis. Numerical tests showed
that the \ppp\ method invariably gave a higher $p$-value (i.e. lower
significance) than the RS
method, and so no data that might have shown a detection with the
rigorous \ppp\ method would have been lost by the selection process. 

The large number of spectra examined means the effects of  multiple trial must be included in the analysis. 
For example, to reach a  global detection significance of only $90.0$\% given a sample of $332$ spectra, we would 
require a single trial significance\footnote{Calculated using the standard
  Bonferroni-type correction factor: $p_1 = 1 - (1-p_N)^{1/N}$, where
  $p_1$ is the single trial $p$-value that gives $p_N$ as the rate of type I
  errors in a set of $N$ independent trials. This sometimes known as
  the \v{S}id\`{a}k equation. In this limit of small $p_N$ and large
  $N$ this tends to $p_1 = p_N/N$. 
} in excess of $99.97$\%. Of the $332$ spectra from $40$ GRBs, $12$ spectra from $10$ GRBs
gave a single trial detection of $\geq 99.9$\% significance in at least one of the methods. 
As the best line candidates in the sample, we now consider each of
these in turn. (All significances are single trial values, unless otherwise stated.) 

\subsection{GRB 050730}

A single Gaussian feature was detected in the spectrum extracted from T+692s to T+792s, which was concurrent with a flare in the WT 
mode data (\citet{Starling_2005} and \citet{Pandey_2006}). An absorbed power law plus a broad Gaussian ($\sigma = 0.34^{+0.08}_{-0.16}$ keV) 
at $1.14^{+0.48}_{-0.44}$ keV provided the best fit to the data with $\chi^{2}/\nu = 47/52$ (table~\ref{table:Spectral_fits}). When the line width 
was restricted to below the detector resolution a Gaussian feature at $0.73^{+0.02}_{-0.03}$ keV was detected ($\chi^{2}/\nu = 57/53$).

\begin{figure}
\begin{center}
     \includegraphics[width=8cm,angle=270]{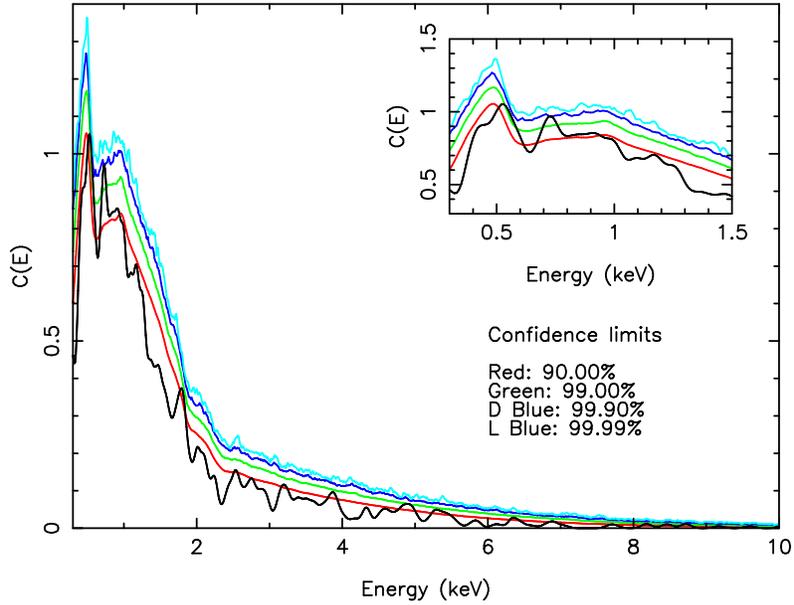}
     \caption{\label{figure:050730_smoothed} GRB 050730 (T+692s to
     T+792s): RS method. Inset focuses on energy range of interest.}
\end{center}
\end{figure}

The Bayes factor was $B_{10}=300$, favoring a line. 
The RS method (fig.~\ref{figure:050730_smoothed}) indicated that a 
line is present in the spectrum at $\sim0.7$ keV with a confidence of 99.90$\%$. This compares favourably to the parameters found in 
the spectral fit when the Gaussian width was restricted to a value below the instrumental resolution. There is no evidence for the broader 
feature found when the width of the Gaussian was a free parameter (see inset to fig.~\ref{figure:050730_smoothed}).

A \ppp\ analysis was carried out in both cases. The significance of the unresolved-width and free-width Gaussian features were found to 
be 88.50$\%$ and 99.92$\%$ respectively. It is surprising that the
\ppp\ analysis appears to favour the wider line at $E_{\rm line} = 1.14^{+0.48}_{-0.44}$ keV,
as there is no evidence of a feature with this energy in the RS plot. However, we note that the large errors on this line energy are 
consistent with a feature at $0.73^{+0.02}_{-0.03}$ keV at the limit of their range.

Applying the \texttt{gain fit} function to this spectrum resulted in an improved fit ($\Delta \chi^{2} = 6$) for an unresolved-width line feature at 
$0.68^{+0.07}_{-0.04}$ keV, with an offset of -55 eV (all other spectral parameters were unchanged within previous limits). Combining this energy offset 
with the error on the line energy is not sufficient to prove an association with the oxygen absorption edge. Applying the \texttt{gain fit} function to 
the free-width Gaussian model was inconclusive, with regards to an association to the oxygen absorption feature, owing to the poorly constrained line 
energy of $< 1.00$ keV. (For further discussion on the application of the \texttt{gain fit} function to this and other GRBs see $\S$7.)

\begin{figure}
\begin{center}
     \includegraphics[width=8cm,angle=270]{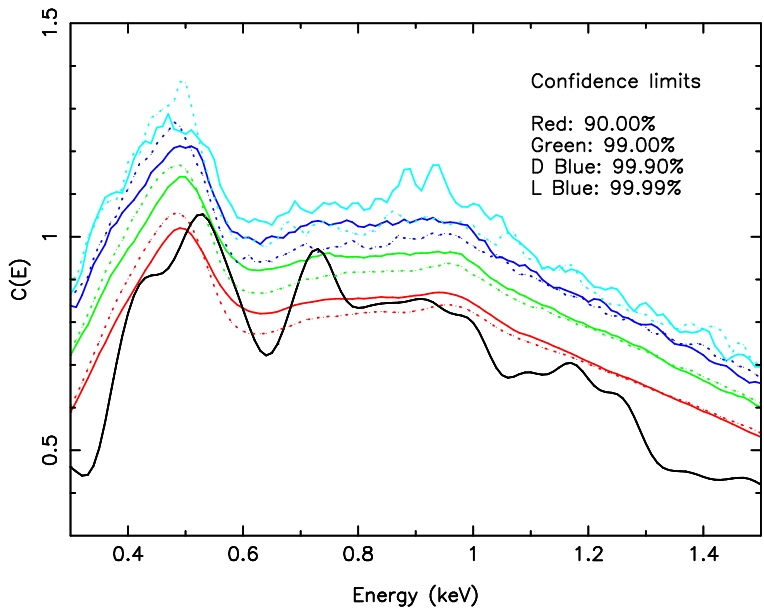}
     \caption{\label{figure:050730_zwabs_smooth} GRB 050730 (T+692s to T+792s): RS comparison between the absorbed power law models containing a single $N_{\rm H}$ component (wabs, dotted lines) and two components (wabs and zwabs, solid lines). Note that the feature becomes far less significant with the addition of the $N_{\rm H}$ column at the appropriate redshift.}
\end{center}
\end{figure}

The redshift for this burst was reported as $z = 3.967$ (\citet{GCN3709}, \citet{GCN3716}, \citet{GCN3732} and \citet{Starling_2005}). Further fits 
were conducted with two $N_{\rm H}$ columns originating from the Galactic column (\texttt{wabs}, fixed at the value given by~\citet{Dickey_1990}) 
and the host galaxy (\texttt{zwabs}). This had the effect of marginally improving the fit for the absorbed power law model ($\chi^{2}/\nu = 61/55$, 
$\Delta \chi^{2} = 3$) with a Galactic column density fixed at $3.21\times10^{20}$ cm$^{-2}$ and a host galaxy component of 
$9.80^{+6.80}_{-6.20}\times10^{21}$ cm$^{-2}$. All of the other spectral parameters were the same as the previous fit within the limits. The fit to 
the other models, containing Gaussian components, did not change
significantly and the parameter values were the same within the error
limits. Bayes factor analysis
including the \texttt{zwabs} component indicated marginal evidence
for line being present ($B_{10} = 5$). Applying the additional \texttt{zwabs} 
component to the RS method (see fig~\ref{figure:050730_zwabs_smooth}) also decreased the significance of the 0.73 keV feature 
from 99.90$\%$ confidence (dotted line) to 99.0$\%$ (solid line). A
\ppp\ analysis, taking the \texttt{zwabs} component into account, found that the 
significance of the free-width feature had decreased to 99.49$\%$ (i.e. $\sim 2.8\sigma$ detection) in this single trial. We conclude that the line detection 
(unresolved or free-width) in GRB 050730 is not significant at 3$\sigma$, and note that the redshift-corrected line energy does not correspond 
to a K-shell transition of a common element.

\subsection{GRB 060109}

This burst had insufficient flux to produce multiple spectra therefore we considered the dataset as a whole. The spectrum covers data
from T+109s to T+199s. An absorbed power law plus a narrow Gaussian at $0.74^{+0.03}_{-0.03}$ keV (width restricted to below the detector 
resolution, $\chi^{2}/\nu = 40/40$) and a free-width Gaussian at $< 0.72$ keV (width = $0.23^{+0.12}_{-0.06}$ keV, $\chi^{2}/\nu = 35/39$) 
were equally good fits to the data (table~\ref{table:Spectral_fits}).

% An absorbed powerlaw plus a narrow Gaussian at $0.74^{+0.03}_{-0.03}$ keV (width restricted to below the detector 
% resolution) provided the best fit to the data with $\chi^{2}/\nu = 40/40$ (table~\ref{table:Spectral_fits}). 

\begin{figure}
\begin{center}
     \includegraphics[width=8cm,angle=270]{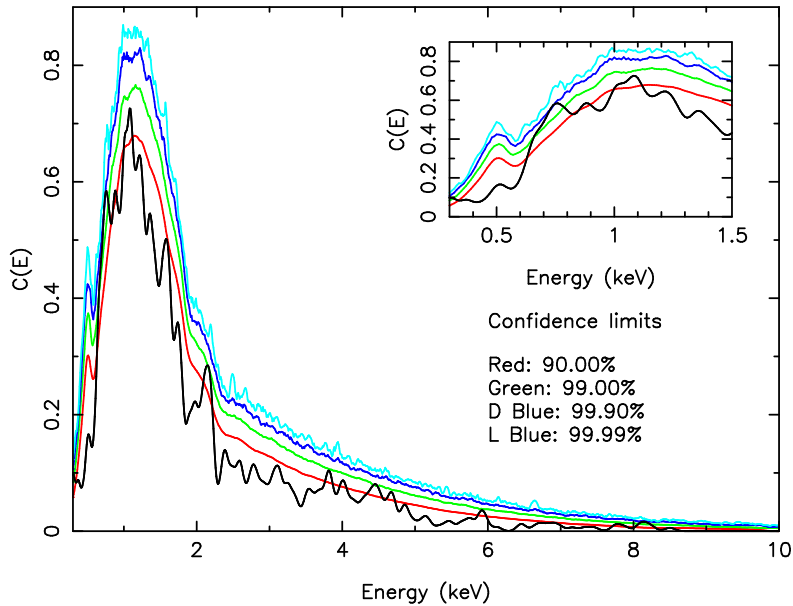}
     \caption{\label{figure:060109_smoothed} GRB 060109: RS results.}
\end{center}
\end{figure}

The Bayes factor for the unresolved-width Gaussian model indicated the presence of a line ($B_{10}=200$), however the same
analysis on the free-width Gaussian was much less convincing
($B_{10}=3$). The RS method indicated that there may be a feature
at $\sim0.7$ keV with a significance of 99.90$\%$ (see fig~\ref{figure:060109_smoothed}). However, the 
\ppp\ method gave only 88.99$\%$ and 99.28$\%$ significance for
unresolved (fixed) and free-width Gaussian lines, respectively. In
$\S$6.1, we showed that the significance of a similar feature
decreased  
below 3$\sigma$ when the spectral fit was changed to include an absorption component at the redshift of the host galaxy. We will show that this is
generally true for those GRBs for which a redshift is known. Unfortunately, in this case, the redshift is not known and we cannot determine whether or not
the same is true.

\subsection{GRB 060111A}

\begin{figure}
\begin{center}
     \includegraphics[width=8cm,angle=270]{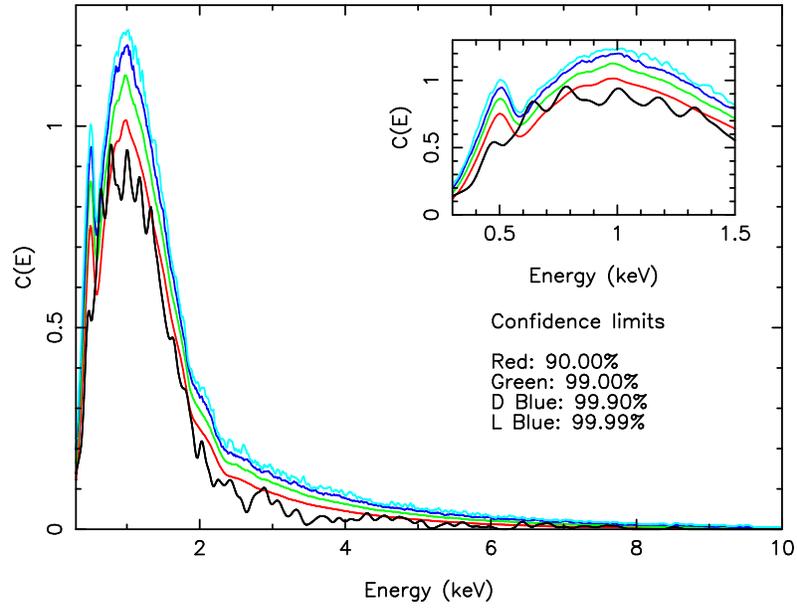}
     \caption{\label{figure:060111A_174_234_smoothed} GRB 060111A
     (T+164s to T+234s): RS results.}
\end{center}
\end{figure}

\begin{figure}
\begin{center}
     \includegraphics[width=8cm,angle=270]{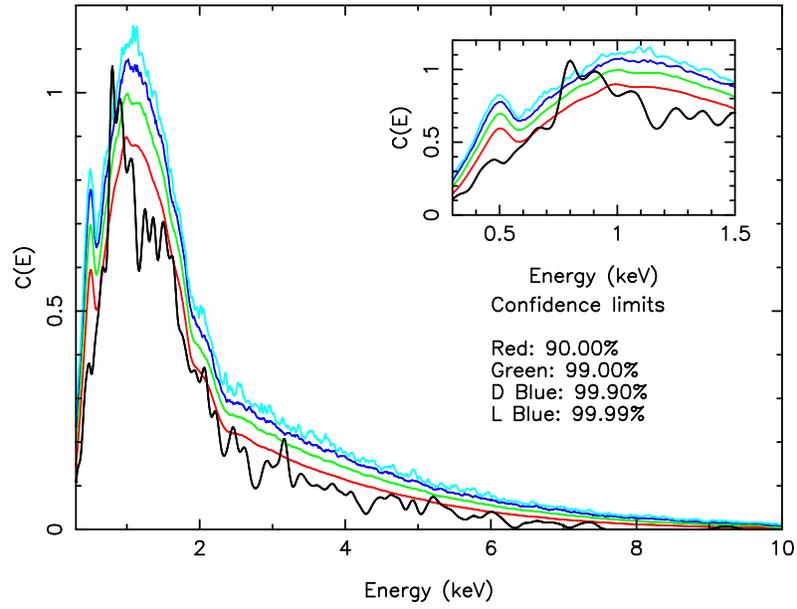}
     \caption{\label{figure:060111A_319_339_smoothed} GRB 060111A
     (T+319s to T+339s): RS results.}
\end{center}
\end{figure}

The data from this burst were split into 11 spectra, covering several flaring events that showed significant spectral variation during the observation.
The Bayes factor ($B_{10}=0.05$) gave no evidence for a free-width line (at $0.65^{+0.09}_{-0.06}$ keV, $W <0.13$ keV)
in the spectrum covering T+174s to T+234s, despite it producing a modest
improvement in the fit ($\Delta \chi^2 = 8$; table~\ref{table:Spectral_fits}). The RS results (fig~\ref{figure:060111A_174_234_smoothed}), 
a feature at $\sim0.65$ keV with $\ge 99.90\%$ confidence. A further feature at $0.79^{+0.02}_{-0.01}$ keV ($W <0.15$ 
keV) was detected in the spectrum covering T+319s to T+339s.
The Bayes factor indicated that the presence of a line in the second spectrum was 
unlikely ($B_{10}=0.1$) but the RS method (fig~\ref{figure:060111A_319_339_smoothed}) suggested an additional spectral feature. 

Whilst the $\sim 0.65$ keV feature for T+174s to T+234s and the $\sim0.79$ keV feature in the T+319s to T+339s both look promising from the 
RS method, the \ppp\ analysis showed that they were only 85.13$\%$ and
99.56$\%$ significant, respectively, not strong detections given the
number of trials (see above).

\subsection{GRB 060115}

\begin{figure}
\begin{center}
     \includegraphics[width=8cm,angle=270]{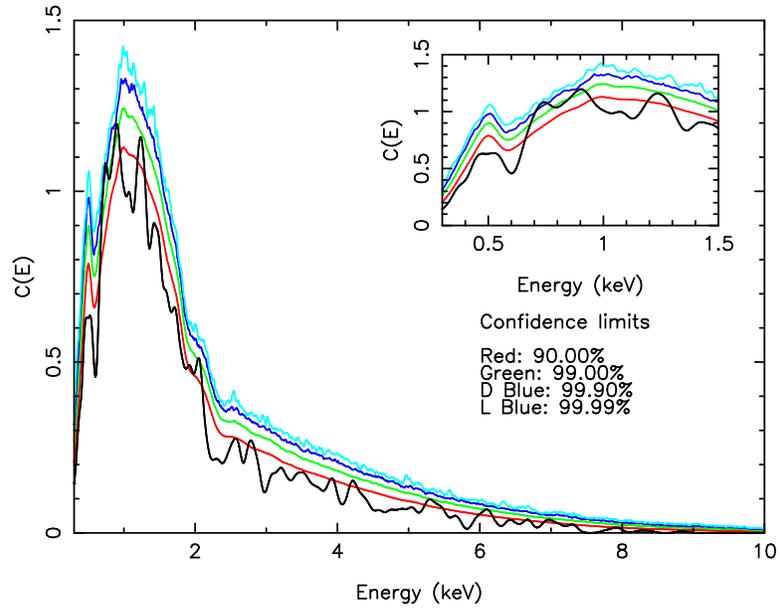}
     \caption{\label{figure:060115_smoothed} GRB 0601115 (T+121s to T+253s): RS results.}
\end{center}
\end{figure}

This burst had insufficient flux to produce multiple spectra; therefore we considered the dataset as a whole. The spectrum covers data from 
T+121s to T+253s. An absorbed power law plus a Gaussian at $0.81^{+0.07}_{-0.07}$ keV with a width of $0.10^{+0.06}_{-0.05}$ keV provided 
the best fit to the data with $\chi^{2}/\nu = 82/77$
(table~\ref{table:Spectral_fits}). The Bayes factor
gave no evidence for a line ($B_{10} = 0.03$), but the RS results
(fig.~\ref{figure:060115_smoothed}) indicated that there was a feature
at $\sim0.75$ keV at the 99.90$\%$  
significance. However, \ppp\ analysis gave only 96.16$\%$
significance.

A redshift of $z = 3.53$ was reported by~\citet{GCN_4520}. Further fits were conducted with two $N_{\rm H}$ columns originating from the Galactic 
column (\texttt{wabs}, fixed at the value given by~\citet{Dickey_1990}) and the host galaxy (\texttt{zwabs}). This led to no change in the statistical fit nor 
parameter values for an absorbed power law model or models containing
Gaussian components. We conclude that the line detection in GRB 060115
is only moderately significance in a single trial, and not 
significant (to 3$\sigma$) in multiple trials, and note that the
redshift-corrected line energy does not correspond to a K-shell
transition of a common element. 

\subsection{GRB 060124}

A precursor $\sim570$ s before the main burst peak allowed {\it Swift's\/} narrow-field instruments to be
positioned on the GRB location $\sim350$ s before the burst occurred~\citep{Romano_2006}. Therefore the WT mode data covered both the prompt emission 
from the burst as well as a portion of the afterglow phase. The flux detected over the observation was sufficient to produce a time series containing 
46 spectra. Bayes factor analysis indicated that eight of these showed
evidence for additional spectral features and a further nine showed evidence 
from the raw $\Delta \chi^2$ improvements. However, RS and \ppp\ analyses carried out on all of these potential 
line spectra revealed only one with an acceptable detection (with both methods giving a significance of $>99.90\%$). This spectrum spanned T+537s to T+542s 
(i.e. occurring just prior to the main burst peak). The best fit model to this spectrum was an absorbed power law plus a broad ($\sigma = 0.48^{+0.17}_{-0.11}$ keV) 
Gaussian component at $2.30^{+0.21}_{-0.23}$ keV (table~\ref{table:Spectral_fits}).

\begin{figure}
\begin{center}
     \includegraphics[width=8cm,angle=270]{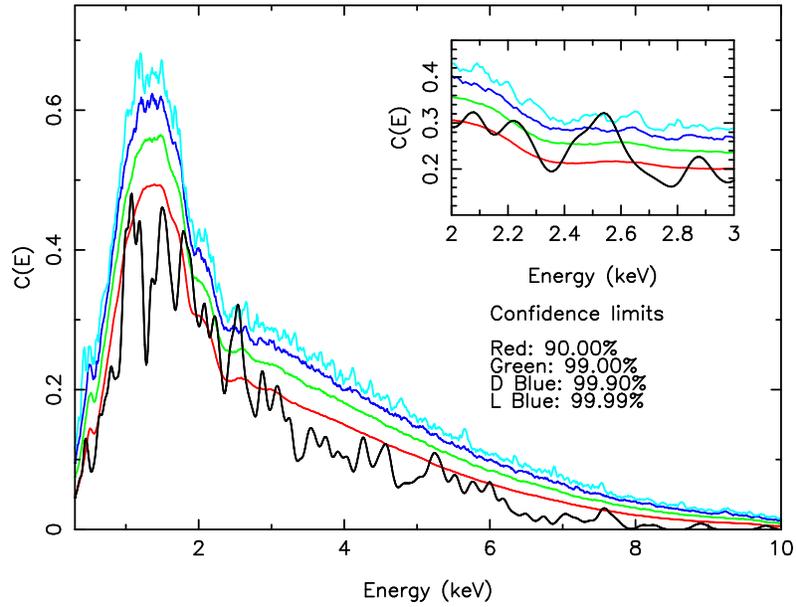}
     \caption{\label{figure:060124_smoothed} GRB 060124 (T+537s to T+542s): RS results.}
\end{center}
\end{figure}

The Bayes factor was $B_{10} = 20$ for a free-width Gaussian feature in this
spectrum. RS results
(fig.~\ref{figure:060124_smoothed}) showed a 99.99$\%$ significance
feature at $\sim2.55$ keV. 
A \ppp\ analysis indicates that the feature is significant to 99.97$\%$.

Whilst this appears to be a significant detection it seems to be very broad for a single line feature, requiring a velocity dispersion of the order
0.5$c$. Using the redshifts of 0.82~\citep{GCN_4591} and 2.297~\citep{GCN_4592} it is possible to identify this feature with $K_{\alpha}$ emission of 
Calcium (4.10 keV) or Cobalt (7.5 keV) respectively. It could in principle be a series of unresolved line features, a thermal component or indicating a 
break in the spectrum. Fitting the spectrum with a blackbody component (kT = $0.76^{+0.14}_{-0.11}$ keV) did not provide a good fit ($\chi^2/\nu = 61/45$) 
nor does an absorbed broken power law model ($\chi^{2}/\nu = 56/45$). 

A further possibility is that it could be due to a poor fit to the gold M-edge as seen in $\S$\ref{section:PKS}. However, applying an energy offset to the data 
does not significantly improve the absorbed power law model ($\Delta \chi^{2}/\nu = 73/45$, offset = -0.08 keV, no change to the slope).

\subsection{GRB 060202}

\begin{figure}
\begin{center}
     \includegraphics[width=8cm,angle=270]{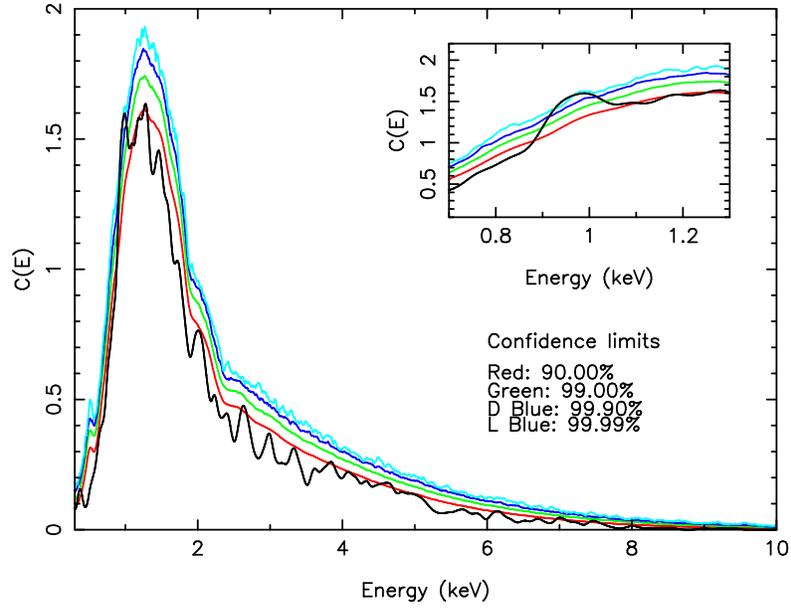}
     \caption{\label{figure:060202_smoothed} GRB 060202 (T+429s to T+529s): RS results.}
\end{center}
\end{figure}

This burst contained sufficient flux to extract 18 spectra. Of these spectra only one, spanning T+429 s to T+529 s, appeared to contain an 
additional spectral feature. The Bayes factor was $B_{10} = 300$ in favor of a single free-width line. An
absorbed power law plus a broad ($< 0.34$ keV) Gaussian feature at $0.94^{+0.05}_{-0.08}$ keV ($\chi^{2}/\nu = 96/100$) was a slightly better 
fit than an absorbed power law alone ($\chi^{2}/\nu = 109/103$; see table~\ref{table:Spectral_fits}).  
RS results for the T+429 s to T+529 s data (fig~\ref{figure:060202_smoothed}) indicated a broad feature at $\sim0.95$ keV, which exceeds the 
99.99$\%$ confidence interval. A \ppp\ analysis of the same data places a significance of 99.74$\%$ on this broad feature. 
No redshift value has been reported for this burst thus we were unable
to perform a well constrained two component absorption fit. 

\subsection{GRB 060210}

\begin{figure}
\begin{center}
     \includegraphics[width=8cm,angle=270]{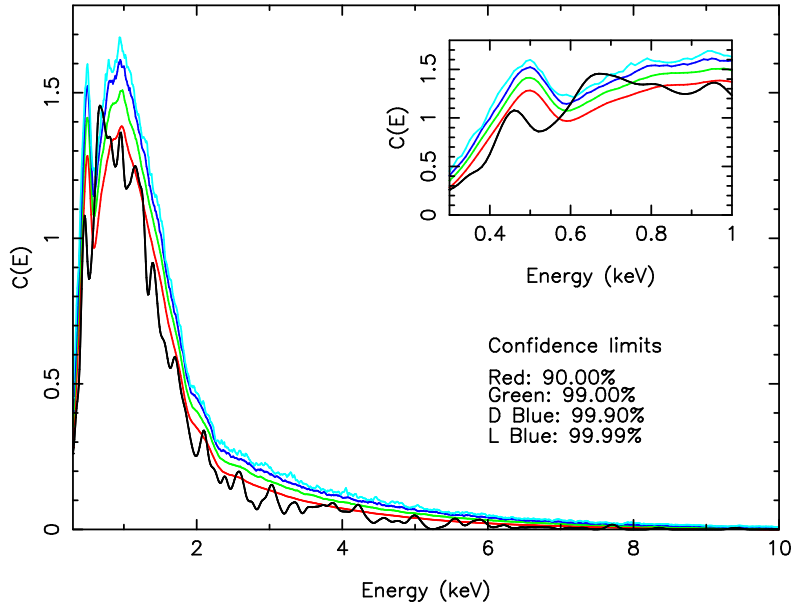}
     \caption{\label{figure:060210_smoothed} GRB 060210 (T+233s to T+353s): RS results.}
\end{center}
\end{figure}

This burst contained sufficient flux to extract a time series containing 8 spectra. Of these spectra only one, spanning T+233s to T+353s, appeared
to contain an additional spectral feature. A model containing a Gaussian feature at 
$0.67^{+0.03}_{-0.04}$ keV (width = $0.06^{+0.05}_{-0.03}$ keV) was a
much better fit than an absorbed power law alone ($\Delta \chi^2 =
13$; table~\ref{table:Spectral_fits}).
The Bayes factor was $B_{10}=1$. The RS results 
(fig~\ref{figure:060210_smoothed}) indicated that a feature 
at $\sim0.65$ keV with a significance of 99.99$\%$. A \ppp\ analysis of the 
spectrum indicated that the same feature is significant to 99.83$\%$. 

A redshift of 3.91 was reported by~\citet{GCN_4729} for this burst. A
two component absorption fit was carried out on the data. This
produced a significantly  improved fit to the absorbed power law ($\Gamma =
2.50^{+0.12}_{-0.11}$) model with a host $N_{\rm H}$ column
contribution of $5.65^{+0.85}_{-0.77}\times10^{22}$  
cm$^{-2}$ and $\chi^{2}/\nu = 80/72$ ($\Delta \chi^{2} = 18$ compared
to the fit with free Galactic absorption only). Adding a Gaussian
component to this gave  $\Gamma = 2.46^{+0.12}_{-0.12}$, a host absorption column of
$5.71^{+0.94}_{-0.83}\times10^{22}$ cm$^{-2}$ and $\chi^{2}/\nu =
76/69$ ($\Delta \chi^{2} = 5$  
compared to the fit with free Galactic absorption only). The addition
of the \texttt{zwabs} component did not change the energy of the
feature but was only able to place  
an upper limit of $< 0.10$ keV on its width. Bayes factor analysis
after allowing for  a \texttt{zwabs} component indicated no evidence
for an additional
spectral feature ($B_{10} = 5 \times 10^{-4}$). 
We can conclude that this feature is most likely a false positive detection.

\subsection{GRB 060218}

\begin{figure}
\begin{center}
     \includegraphics[width=8cm,angle=270]{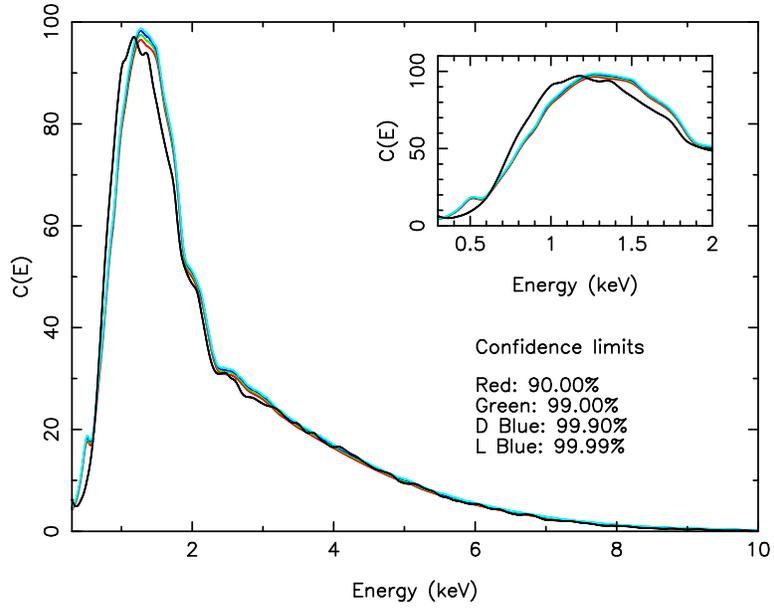}
     \caption{\label{figure:060218_all_smoothed} GRB 060218 (T+159s to
     T+2770s): RS results.}
\end{center}
\end{figure}

\begin{figure}
\begin{center}
     \includegraphics[width=8cm,angle=270]{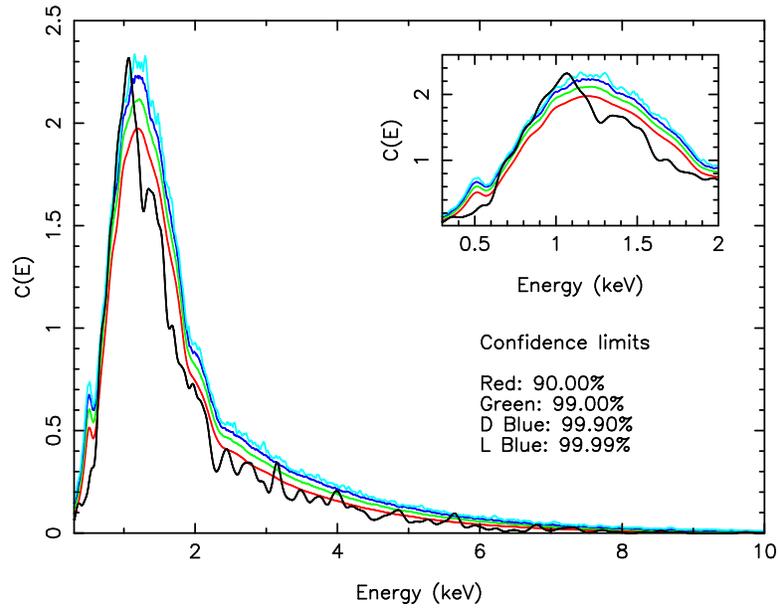}
     \caption{\label{figure:060218_section_smoothed} GRB 060218
     (T+2359s to T+2409s): RS results.}
\end{center}
\end{figure}

~\citet{Campana_2006} have reported on the association of this burst with SN2006aj and the presence of a thermal component in the X-ray spectrum in 
great detail. Our analysis concurs with their results. 
The data were split into $53$ time intervals, from which 
the Bayes factor analysis indicated
additional component in the spectrum in all data from $\sim$T+750 s
(with $B_{10} > 50$). This was confirmed by RS and \ppp\ analysis.
The RS results (fig~\ref{figure:060218_all_smoothed}, T+159 s to T+2770 s) indicated that this feature is unlikely to be a Gaussian emission line 
as its profile was too broad. It is possible that it could be a series
of unresolved lines, however, a power law plus blackbody 
component gave the best fit to all of the spectra suggesting an
additional spectral feature. Similarly, individual time-slices (see
fig~\ref{figure:060218_section_smoothed}, 
T+ 2359 s to T+2409 s, for one such example) show the presence of this
broad feature, which appears to evolve over
time~\citep{Campana_2006}. 

\subsection{GRB 060418}

A time series of 12 spectra were extracted from this GRB, two of which
appear to contain additional spectral components. 
These were the spectra spanning T+119 s to T+129 s and T+169 s to T+194 s.

\begin{figure}
\begin{center}
     \includegraphics[width=8cm,angle=270]{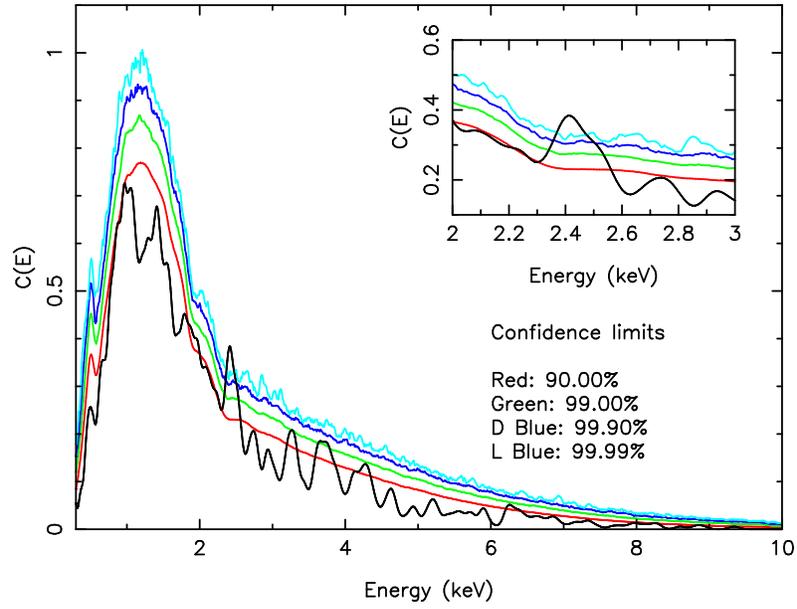}
     \caption{\label{figure:060418_119_129_smoothed} GRB 060418
     (T+119s to T+129s): RS results.}
\end{center}
\end{figure}

A Gaussian component at $2.42^{+0.02}_{-0.03}$ keV improved the fit to
the T+119 s to T+169 s data by $\Delta \chi^2 = 16$ (see table~\ref{table:Spectral_fits}), although 
the Bayes factor was unconvincing ($B_{10} = 0.05$).
The RS analysis (fig.~\ref{figure:060418_119_129_smoothed}) showed a feature at this 
energy that clearly exceeded the 99.99$\%$ confidence limit. A \ppp\
analysis found 99.85$\%$ significance for the same feature. However, as noted
previously in the analysis for GRB 060124 and PKS0745-19
($\S$\ref{section:PKS}), a feature at this energy is coincident with
the gold M-edge. 

\begin{figure}
\begin{center}
     \includegraphics[width=8cm,angle=270]{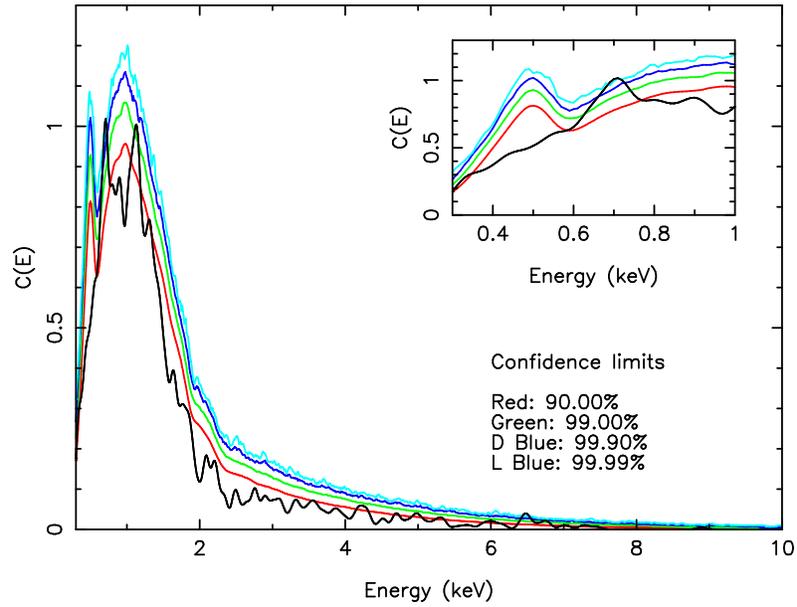}
     \caption{\label{figure:060418_169_194_smoothed} GRB 060418
     (T+169s to T+194s): RS results.}
\end{center}
\end{figure}

A similar improvement in the fit was found for 
the second spectrum (T+169 s to T+194 s), with $\Delta \chi^2 = 10$,
the Bayes factor was more promising ($B_{10} = 30$).
An unresolved-width Gaussian at $< 0.75$ keV provided
the best fit to this spectrum with $\chi^{2}/\nu = 43/49$
(table~\ref{table:Spectral_fits}). RS
(fig.~\ref{figure:060418_169_194_smoothed}) and \ppp\ analysis 
supported the presence of this feature at  
the 99.99$\%$ and 99.98$\%$ confidence limit respectively. 

The 2.42 keV feature of the T+119 s to T+129 s spectrum can be explained by the gold M-edge but the 0.69 keV feature of the T+169 s to T+194 s spectrum 
cannot be matched to another elemental absorption edge in the same manner. Two component absorption fits were carried out with a $N_{\rm H}$ column 
density of $9.17\times10^{20}$ cm$^{-2}$ from our Galaxy and a contribution from the host galaxy at $z = 1.49$ (\citet{GCN_4969} and \citet{GCN_4974}). 
This produced a significant improvement in the absorbed power law model fit, which gave $\Gamma = 2.48^{+0.18}_{-0.13}$ and a host 
$N_{\rm H} = 0.73^{+0.23}_{-0.20} \times10^{22}$ cm$^{-2}$ ($\chi^{2}/\nu = 53/52$, $\Delta \chi^{2} = 9$ compared to the fit with free Galactic 
absorption only). The addition of a \texttt{zwabs} component to the Gaussian model gave $\Gamma = 2.22^{+0.22}_{-0.15}$, a line with an energy of $< 0.65$ keV 
and width of $0.47^{+0.06}_{-0.08}$ keV and a host absorption $<
0.21\times10^{22}$ cm$^{-2}$ ($\chi^{2}/\nu = 46/49$). The Bayes factor for the spectra 
containing the \texttt{zwabs} component indicates that the odds of an
additional spectral component have been significantly reduced to
$B_{10}=1$. We can conclude that  
this feature is most likely not real, but a spurious detection due to the baseline assumption of no host galaxy absorption.

\subsection{GRB 060428B}

\begin{figure}
\begin{center}
     \includegraphics[width=8cm,angle=270]{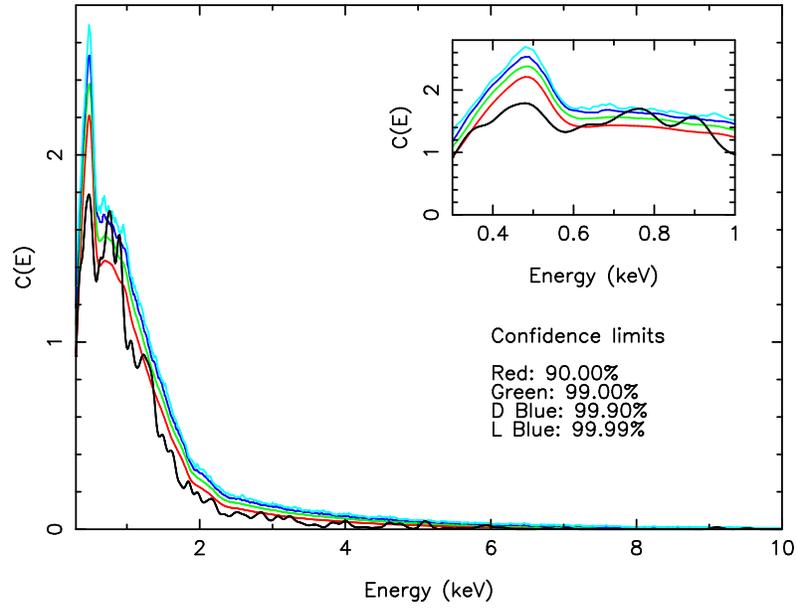}
     \caption{\label{figure:060428B_212_252_smoothed} GRB 060428B
     (T+212s to T+252s): RS results.}
\end{center}
\end{figure}

Data from this burst were split into two sets, T+212 s to T+252 s and T+252 s to T+418 s. An absorbed power law model was a poor fit
to the first spectrum with $\chi^{2}/\nu = 78/63$ whilst an absorbed power law plus a Gaussian feature at $0.76^{+0.05}_{-0.06}$ keV 
(width = $0.09^{+0.05}_{-0.03}$ keV) was a much better fit with $\chi^{2}/\nu = 63/60$. However, the Bayes factor was less encouraging with
$B_{10}=0.1$ against a line feature. The RS analysis (fig.~\ref{figure:060428B_212_252_smoothed}) indicated the presence of two 
possible features; one at $\sim0.75$ keV at a significance of
99.99$\%$ and another at $\sim0.90$ keV at 99.90$\%$. 
However, no stable spectral fit could be found using an emission line
at $\sim0.9$ keV, hence it was not possible to calculate a Bayes
factor, nor calculate the $\Delta \chi^2$ needed for a \ppp\ calculation.
A \ppp\ analysis of the feature at $0.76^{+0.05}_{-0.06}$ keV yielded a significance of 99.85$\%$.

\begin{figure}
\begin{center}
     \includegraphics[width=8cm,angle=270]{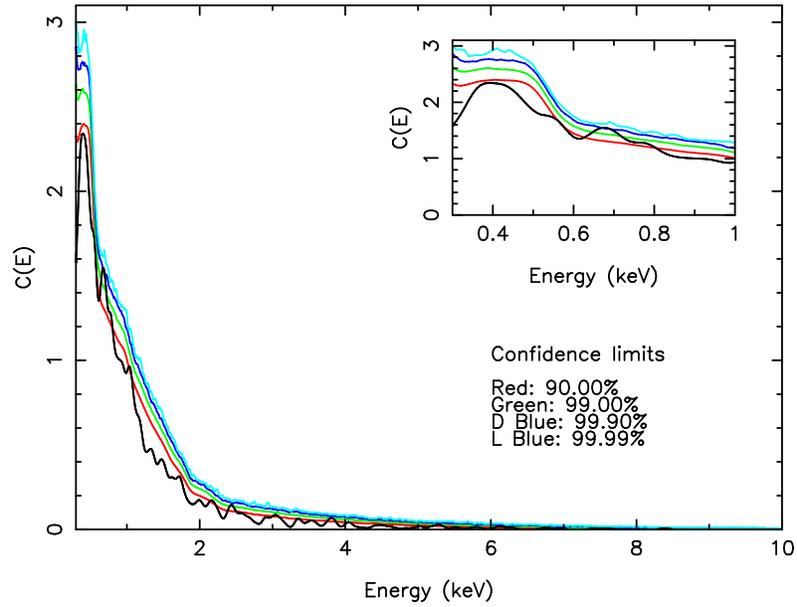}
     \caption{\label{figure:060428B_252_418_smoothed} GRB 060428B
     (T+252s to T+418s): RS results.}
\end{center}
\end{figure}

The second spectrum, T+252 s to T+418 s, was best fitted by an absorbed power law model ($\chi^{2}/\nu = 58/64$, table~\ref{table:Spectral_fits}) 
and the Bayes factor gave only very weak evidence to indicate a line ($B_{10}=3$).
RS analysis (fig~\ref{figure:060428B_252_418_smoothed}) indicated a possible feature at $\sim0.7$ keV with a significance of
99.90$\%$, however a \ppp\ analysis placed a much lower significance of 95.87$\%$ on this.

No redshift value has been reported for this burst, preventing us from performing a constrained two component absorption fit. This could potentially 
determine if the features at $\sim 0.7$ keV are due to poor modeling of the absorption continuum due to not including a component from the host
galaxy. 

\subsection{Use of alternative prior}
\label{section:alt-prior}

In section~\ref{section:priors} we discussed two different choices for
assigning an uninformative prior to the line normalization. The
approximate
Bayes factors given above were calculated assuming a uniform prior for
the line normalization, but
using the Jeffreys prior did not change the results significantly.
For example, GRB 060115 changed from $B_{10}=0.03$ with the
uniform prior to $B_{10}=0.05$ with the Jeffrey's prior.
At the other extreme, the favorable Bayes factor of $B_{10}=300$ for GRB
060202 (T+[429-529] s) using the uniform prior was virtually unchanged.
Spectra that were not included for further analysis, due to a low
Bayes factor ($B_{10} < 1.0$) and a small $\chi^2$ improvement
($\Delta \chi^2 < 4.61$) were similarly affected by a change in
priors. In general the Bayes factors changed very little between
uniform and Jeffreys priors, reflecting the fact that typical best-fitting
line normalizations were usually
$\sim 10$\% of the
total flux (see section~\ref{section:priors}).

\section{Discussion of {\it Swift} XRT results}\label{section:disco}

\begin{figure}
\begin{center}
     \includegraphics[width=8.5cm,angle=270]{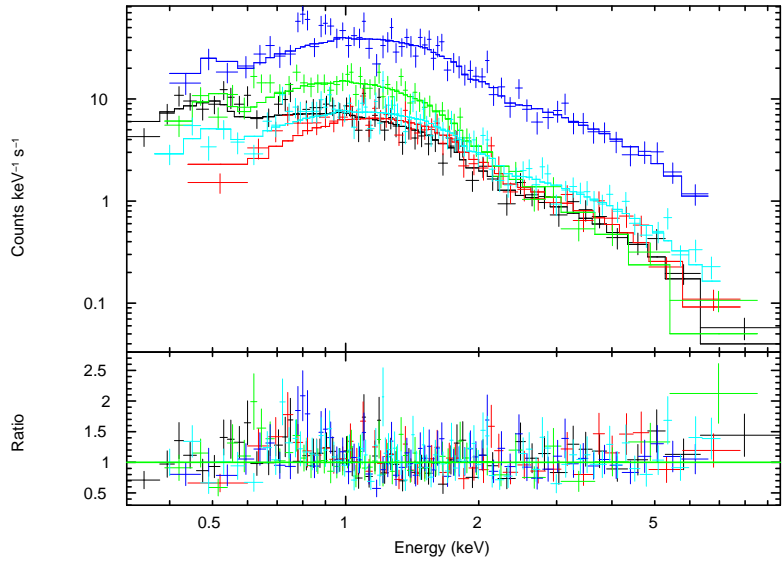}
     \caption{\label{figure:050730_060115} Spectra modeled with an absorbed power law model: GRB 050730 (black), GRB 060109 (red), GRB 060111A (T+174 s to T+234 s, green), 060111A (T+319 s to T+339 s, dark blue) and 060105 (light blue). Note the residuals around 0.7 keV.}
\end{center}
\end{figure}

\begin{figure}
\begin{center}
     \includegraphics[width=8.5cm,angle=270]{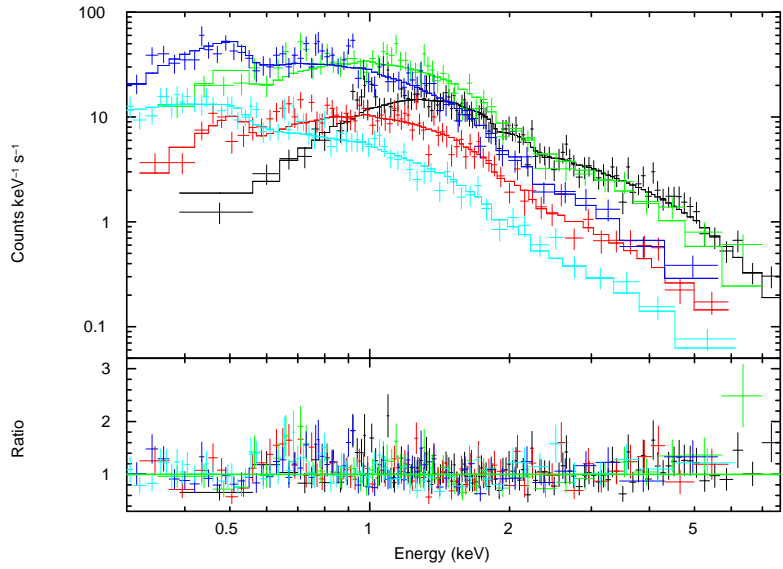}
     \caption{\label{figure:060202_060428B} Spectra modeled with an absorbed power law model: GRB 060202 (black), GRB 060210 (red), GRB 060418 (T+169 s to T+194 s, green), 060428B (T+212 s to T+252 s, dark blue) and 060428B (T+252 s to T+418 s, light blue). Note the residuals around 0.7 keV.}
\end{center}
\end{figure}

\begin{figure}
\begin{center}
     \includegraphics[width=8.5cm,angle=270]{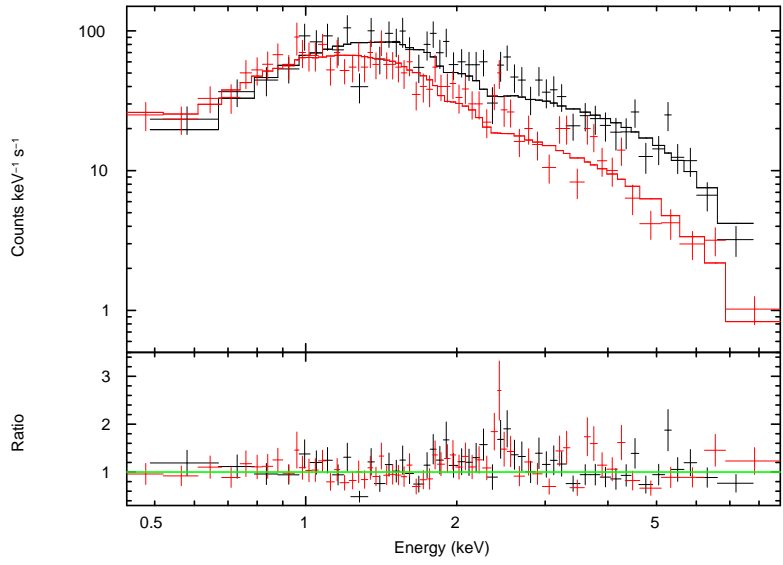}
     \caption{\label{figure:2.3_features} Spectra modeled with an absorbed power law model: GRB 060124 (black) and GRB 060210 (T+119 s to T+129 s, red). Note the residuals around 2.3 keV.}
\end{center}
\end{figure}

The previous section shows that of 332 WT mode spectra analyzed by our
methods only 12 produced possible detections at $\ge$ 99.90\% (single
trial). These detections were tightly clustered around two energies in the
observer frame: 0.64-0.94 kev (10/12, figs~\ref{figure:050730_060115}  
and~\ref{figure:060202_060428B}) and 2.30-2.49 keV (2/12,
fig~\ref{figure:2.3_features}), with equivalent widths of $\sim$0.9
keV and $\sim$0.5 keV respectively.

The coincidence of many spectral feature detections close to 0.7 keV is suspicious as we would expect intrinsic GRB emission line features to be located
at different {\it observed\/} energies, as the GRBs span a large range of redshifts. This clustering strongly hints at an instrumental origin.
Modeling the WT mode spectra with an energy offset, in case of imperfect bias subtraction at the processing stage, improved the fit statistics for an absorbed 
power law model (average $\Delta \chi^{2} \sim 5$). However, even if the combined error on the line energy and the offset corrections (1 to 70 eV) are taken, 
this would still not be enough to provide a plausible association with the oxygen K-edge as seen in the PKS 0745-19 example ($\S$4).

An additional absorption component, at the host galaxy redshift, was applied to those candidates with a known redshift measurement. In every case the
feature at 0.7 keV became insignificant and we expect that the same reduction in significance would occur if we were able to conduct well constrained
two absorption component fits to the GRBs with unknown redshifts. It could be argued that this decrease in line significance stems only from the increased 
complexity in the model. If this were the case we should see an overall increase in the level of the RS contours over the whole energy range. In 
fig~\ref{figure:050730_zwabs_smooth} we can see that the effect of the adding a \texttt{zwabs} component is not uniform across the energy range. It has 
negligible effect at energies $>$1.2 keV. Below this energy the \texttt{zwabs} component acts to increase the total absorption at very low energies
($<$0.55 keV) and decrease it in the 0.55 keV to 1.2 keV range. Thus the addition of the second absorption component is imposing a real, energy dependant effect 
on the confidence contours, rather than increasing them uniformly across the whole energy range. We conclude that the absorption is not being modeled accurately 
at low energies by assuming that all of the $N_{\rm H}$ column is at a redshift of zero.

We have confirmed that the features at 2.3 keV are not due to bad pixel or hot column issues. The two detections have the following single-trial significances:
99.97\% (GRB 060124) and 99.85\% (GRB 060418). Taken in the context of all the trials performed these significances become 90.09\% ($<1.7\sigma$) and
50.20\% ($<0.7\sigma$) respectively. There is no significant impovement to the model fits if the \texttt{gain fit} function is applied. Adding a blackbody 
component to the underlying absorbed power law or using an absorbed broken power law does not yield a significantly improved fit either. We note, however, that 
both of these features are coincident with the gold M-edge complex.

Since all of the features are found to be narrowly clustered around two energies in the observer frame, one of which was also found in the PKS 0745-19 example,
it is our conclusion that none are real detections of emission lines in GRB spectra, but are instead either due to residual calibration issues, imperfect bias 
subtraction at the processing stage or incorrect modeling of the host galaxy absorption column (most likely for the 0.7 keV features).

In addition we would expect of the order of 3 false positive detections at 99.00\% for 332 spectral analyses.

It is also interesting that the majority of GRBs with potential features occur in the first few months of 2006. There are no physical, instrumental
or calibration issues associated with that period of operations that could explain such temporal clustering. However, we note that the actual response of 
the XRT CCD is possibly evolving while the response is modeled by the calibration files as being constant.

~\citet{Butler_2007} has recently published analysis citing the detection of line complexes in GRBs 050714B, 050822, 060202 and 060218. However, the same paper 
and \citet{Butler_Kocevski_2007} also provided alternate reasons for these apparent lines detections (thermal components or broken power laws). GRB 050714B
was not included in our selection of bursts as the WT mode spectrum did not contain sufficient counts to meet our minimum criteria. 

We found no compelling evidence from GRB 050822 using any of our analysis methods to suggest that there were any line features in these data, hence it does not 
feature in $\S$\ref{section:Swift}. \citet{Butler_2007} quotes a
significance of 4.4$\sigma$ for a complex of 5 lines (0.81, 0.91,
1.04, 1.23 and 1.49 keV) for one spectrum  
spanning T+489.5 s to T+509.4 s, with equivalent widths of 82, 142, 194, 221 and 265 eV respectively. We note that the features are spaced 100 eV apart,  
though no obvious physical explanation for this presents itself. It should be noted that Butler's spectra ($\sim500$ counts) is contained within our 
analysis from  T+471 s to T+661 s ($\sim860$ counts); our spectrum cannot be sub-divided further whilst still being directly comparable to the rest of 
the data analysis presented in this paper. None of the power law (or cutoff power law) models tested provide a good fit to the data (table~\ref{table:050822}), though 
cutoff power laws appear to give a much better fit to the data. Adding a blackbody component to an absorbed power law gave poorly constrained parameters: 
$\Gamma < 1.98$; kT $= 0.17^{+0.02}_{-0.02}$; and $N_{\rm H}$ $= 4.58^{+5.16}_{-3.08}\times10^{20}$ cm$^{-2}$ ($\chi^{2}/\nu = 46/38$).

RS analysis of our spectra compared to a base model of an absorbed cutoff power law shows a feature at $\sim 0.7$ keV at $> 99.99 \%$ 
significance. However, \ppp\ analysis only placed a significance of 99.74$\%$ on the same feature. In addition the appearance of a feature at 0.7 keV 
indicates that the absorption column may not have been modeled accurately, as seen in other bursts in the previous section. As there is no report of a 
redshift for this burst we cannot confirm this by carrying out a well constrained two component absorption fit.

060202 and 060218 have both been discussed in the previous section. We agree with the presence of a blackbody component in 060218 \citep{Campana_2006}. 
The single feature found in 060202 only occurred in 1 of the 18 time sliced spectra with a significance of 99.74$\%$ (single trial). If we consider the 
multiple trials carried out this drops to a significance of $95.4$\%. Our analysis indicates that it is a broad feature ($\sigma < 0.34$ keV) at 0.9 keV 
rather than a series of resolved or overlapping lines, however it does occur at the same energy over which~\citet{Butler_2007} reports 4 individual narrow 
lines. We find no evidence for the reported feature at $4.70\pm0.07$ keV.

\section{Conclusions} \label{section:conc}

Analysis of the galaxy cluster PKS 0745-19, which has a known 6.07 keV emission line, produced a
convincing detection by all methods and also uncovered two further features at 0.6 keV and 2.3 keV.
Both features are likely to be due to an energy scale offset that causes the instrumental oxygen and
gold absorption edges respectively to be poorly fitted ($\S$~\ref{section:PKS}). A series of
simulations over a range of emission line parameters has allowed us to estimate the sensitivity to
Gaussian-like features, both broad and narrow. For all three methods, using GRB parameters typical for
{\it Swift\/} bursts, the optimum range for emission line detection was found to be $0.4 - 6$ keV,
with line equivalent widths as low as $\sim 50$~eV detectable in principle from data with only $\sim
1600$ counts.

Of the 332 WT mode spectra from real GRBs, only 12 produced possible detections at $\ge$99.90\%
(single trial). These were all located around two energies in the observer frame: 0.7 keV (10/12) and
2.3 keV (2/12). The coincidence of many spectral features close to 0.7 keV is suspicious as we would
expect intrinsic GRB emission line features to be located at different {\it observed\/} energies, as
the GRBs span a large range of redshifts. For those candidates with a redshift measurement the feature
at 0.7 keV becomes insignificant once an absorption component at the redshift of the host galaxy is
applied. We expect that the same reduction in significance would occur if we were able to apply well
constrained two absorption component fits to the GRBs with unknown redshifts. The 2.3 keV features are
thought to be associated with the gold M-edge.

Since all of the features are found narrowly clustered around two energies, one of which was also
found in the PKS 0745-19 spectrum (2.3 keV), it is our conclusion that all of these features are
either due to calibration issues, imperfect bias subtraction at the processing stage or incorrect
modeling of the host absorption column (most likely case for the 0.7 keV features), rather than GRB
emission line detections. The only non-power law emission component we accept as intrinsic is the
blackbody component detected in GRB 060218.

\acknowledgments

{\it We wish to thank Evert Rol for useful discussions and an
  anonymous referee for very detailed comments that prompted us to
  clarify the discussion of the detection methods.
  This work is supported at the University of Leicester by the Particle Physics and Astronomy Research Council (PPARC), at PSU by NASA and in Italy by funding from ASI. CPH gratefully acknowledges support from a PPARC studentship. SV, JPO, ER, KLP, AN, OG, MRG, PE and RS acknowledge PPARC support. This research has made use of data obtained through the UK Swift Data Archive, provided by the University of Leicester. \/}

% ============================
% ============================

 % ============================
% ============================

% ============================
% ============================

% ============================
% ============================

% ============================
% ============================

% ============================
% ============================

\clearpage

% ============================
\begin{deluxetable}{crrrrrrr}
\tablecaption{Summary of spectral fits for all candidate spectra.}
\tabletypesize{\footnotesize}
\tablecolumns{8}
\tablehead{
\colhead{Model$^{a}$} &
\colhead{Photon} &
\colhead{Line energy} & 
\colhead{Line width} &
\colhead{Line norm. ($\times10^{-2}$} &
\colhead{Equiv. width} &
\colhead{$N_{\rm H}$} &
\colhead{$\chi^{2}/\nu$} 
\\
\colhead{} &
\colhead{index} &
\colhead{(keV)} &
\colhead{(keV)} &
\colhead{photons cm$^{-2}$ s$^{-1}$)} &
\colhead{(eV)} &
\colhead{($\times10^{20}$ cm$^{-2}$)} &
\colhead{} 
\\
}
\startdata
\multicolumn{8}{l}{GRB 050730 (T+692s to T+792s)} \\
1   & $2.03^{+0.14}_{-0.13}$  & ...                    & ...                    & ...                    & ... & $5.84^{+2.32}_{-2.14}$ & 64/55  \\
2   & $1.98^{+0.15}_{-0.13}$ & $0.73^{+0.02}_{-0.03}$  & $<$ Inst. res.         & $0.95^{+0.58}_{-0.57}$ & 50 & $8.88^{+6.57}_{-4.72}$ & 57/53   \\
3   & $1.78^{+0.18}_{-0.19}$  & $1.14^{+0.48}_{-0.44}$ & $0.34^{+0.08}_{-0.16}$ & $22.5^{+7.5}_{-18.5}$   & 3400 & $7.08^{+3.95}_{-4.03}$ & 47/52 \\
 & & & & & & \\
\multicolumn{8}{l}{GRB 060109 (T+109s to T+199s)} \\
1   & $2.29^{+0.18}_{-0.17}$  & ...                    & ...                    & ...                    & ... & $31.9^{+5.5}_{-5.0}$ & 48/42  \\
2   & $2.29^{+0.20}_{-0.17}$  & $0.74^{+0.03}_{-0.03}$ & $<$ Inst. res.         & $3.11^{+3.08}_{-1.95}$ & 94 & $34.4^{+6.4}_{-5.6}$ & 40/40  \\
3   & $2.20^{+0.18}_{-0.18}$  & $< 0.72$               & $0.23^{+0.12}_{-0.06}$ & $17.5^{+13.8}_{-7.7}$  & 560 & $39.6^{+9.8}_{-7.6}$ & 35/39 \\
 & & & & & & \\
\multicolumn{8}{l}{GRB 060111A (T+174s to T+234s)} \\
1   & $3.05^{+0.22}_{-0.20}$  & ...                    & ...                    & ...                    & ... & $29.9^{+4.8}_{-4.5}$ & 52/50  \\
2   & $3.09^{+0.23}_{-0.21}$  & $0.64^{+0.03}_{-0.03}$ & $<$ Inst. res.         & $11.6^{+15.2}_{-4.9}$  & 73 & $32.6^{+5.7}_{-5.1}$ & 44/48  \\
3   & $3.07^{+0.09}_{-0.21}$  & $0.65^{+0.09}_{-0.06}$ & $< 0.13$               & $14.0^{+16.0}_{-10.1}$ & 94 & $33.0^{+7.9}_{-5.8}$ & 44/47 \\
 & & & & & & \\
\multicolumn{8}{l}{GRB 060111A (T+319s to T+339s)} \\
1   & $1.97^{+0.14}_{-0.14}$  & ...                    & ...                    & ...                    & ... & $18.8^{+3.9}_{-3.6}$ & 69/61  \\
2   & $1.94^{+0.14}_{-0.14}$  & $0.79^{+0.02}_{-0.01}$ & $<$ Inst. res.         & $9.42^{+4.86}_{-4.18}$ & 80 & $19.5^{+4.1}_{-3.9}$ & 54/59 \\
3   & $1.94^{+0.07}_{-0.09}$  & $0.79^{+0.02}_{-0.01}$ & $< 0.15$ & $9.42^{+11.9}_{-4.18}$               & 80 & $19.5^{+2.7}_{-2.2}$ & 54/58 \\
 & & & & & & \\
\multicolumn{8}{l}{GRB 060115 (T+121s to T+253s)} \\
1   & $1.88^{+0.12}_{-0.11}$  & ...                    & ...                    & ...                    & ... & $16.6^{+3.3}_{-3.1}$ & 93/80  \\
2   & $1.85^{+0.12}_{-0.12}$  & $0.89^{+0.03}_{-0.03}$ & $<$ Inst. res.         & $0.67^{+0.44}_{-0.42}$ & 39 & $16.3^{+3.1}_{-3.2}$ & 86/78  \\
3   & $1.82^{+0.13}_{-0.11}$  & $0.81^{+0.07}_{-0.07}$ & $0.10^{+0.06}_{-0.05}$ & $2.09^{+2.21}_{-1.22}$ & 100 & $17.0^{+4.3}_{-2.7}$ & 82/77 \\
& & & & & & \\
\multicolumn{8}{l}{GRB 060124 (T+537s to T+542s)} \\
1   & $1.30^{+0.16}_{-0.14}$  & ...                    & ...                    & ...                    & ... & $29.5^{+6.7}_{-8.0}$ & 72/47  \\
2   & $1.29^{+0.15}_{-0.15}$  & $2.49^{+0.06}_{-0.01}$ & $<$ Inst. res.         & $11.8^{+6.3}_{-6.5}$   & 800 & $27.8^{+9.3}_{-7.8}$ & 62/45  \\
3   & $1.13^{+0.19}_{-0.24}$  & $2.30^{+0.21}_{-0.23}$ & $0.48^{+0.17}_{-0.11}$ & $57.7^{+42.0}_{-25.2}$ & 150 & $18.3^{+9.7}_{-9.9}$ & 51/44 \\
& & & & & & \\
\multicolumn{8}{l}{GRB 060202 (T+429s to T+529s)} \\
1   & $2.16^{+0.11}_{-0.10}$  & ...                    & ...                    & ...                    & ... & $47.1^{+4.3}_{-4.0}$ & 109/103  \\
2   & $2.15^{+0.11}_{-0.11}$  & $0.94^{+0.03}_{-0.02}$ & $<$ Inst. res.         & $2.69^{+1.51}_{-1.30}$ & 54 & $48.1^{+4.8}_{-4.2}$ & 97/101  \\
3   & $2.12^{+0.10}_{-0.12}$  & $0.94^{+0.05}_{-0.08}$ & $< 0.34$               & $4.94^{+5.01}_{-2.80}$ & 99 & $50.0^{+1.1}_{-5.0}$ & 96/100 \\
& & & & & & \\
\multicolumn{8}{l}{GRB 060210 (T+233s to T+353s)} \\
1   & $2.72^{+0.16}_{-0.15}$  & ...                    & ...                    & ...                    & ... & $20.6^{+3.1}_{-2.9}$ & 98/72  \\
2   & $2.71^{+0.16}_{-0.15}$  & $0.66^{+0.04}_{-0.02}$ & $<$ Inst. res.         & $4.63^{+3.06}_{-2.40}$ & 63 & $21.5^{+3.5}_{-3.1}$ & 85/70  \\
3   & $2.68^{+0.18}_{-0.16}$  & $0.67^{+0.03}_{-0.04}$ & $0.06^{+0.05}_{-0.03}$ & $7.06^{+5.94}_{-4.54}$ & 100 & $21.6^{+4.0}_{-3.2}$ & 81/69  \\
 & & & & & & \\
\multicolumn{8}{l}{GRB 060418 (T+119s to T+129s)} \\
1   & $1.82^{+0.13}_{-0.12}$  & ...                    & ...                    & ...                    & ... & $24.3^{+5.1}_{-4.6}$ & 72/59  \\
2   & $1.82^{+0.13}_{-0.12}$  & $2.42^{+0.02}_{-0.03}$ & $<$ Inst. res.         & $7.21^{+3.00}_{-2.97}$ & 190 & $23.7^{+5.0}_{-4.6}$ & 56/57  \\
3   & $1.82^{+0.13}_{-0.12}$  & $2.42^{+0.02}_{-0.04}$ & $< 0.14$               & $7.21^{+4.74}_{-2.98}$ & 190 & $23.7^{+4.6}_{-4.5}$ & 56/56 \\
& & & & & & \\
\multicolumn{8}{l}{GRB 060418 (T+169s to T+194s)} \\
1   & $2.70^{+0.22}_{-0.19}$  & ...                    & ...                    & ...                    & ... & $22.3^{+4.3}_{-3.9}$ & 62/52  \\
2   & $2.67^{+0.07}_{-0.12}$  & $0.69^{+0.02}_{-0.02}$ & $<$ Inst. res.         & $10.8^{+5.6}_{-5.5}$   & 58 & $22.2^{+2.1}_{-2.0}$ & 52/50  \\
3   & $1.82^{+0.22}_{-0.66}$  & $< 0.75$               & $0.57^{+0.09}_{-0.20}$ & $110^{+31}_{-75}$ & 2300 & $11.0^{+0.1}_{-0.1}$ & 43/49 \\
& & & & & & \\
\multicolumn{8}{l}{GRB 060428B (T+212s to T+252s)} \\
1   & $3.02^{+0.18}_{-0.16}$  & ...                    & ...                    & ...                    & ... & $11.6^{+2.6}_{-2.3}$ & 78/63  \\
2   & $2.94^{+0.18}_{-0.16}$  & $0.77^{+0.03}_{-0.02}$ & $<$ Inst. res.         & $3.56^{+1.84}_{-1.87}$ & 39 & $10.6^{+2.6}_{-2.2}$ & 69/61  \\
3   & $2.83^{+0.16}_{-0.16}$  & $0.76^{+0.05}_{-0.06}$ & $0.09^{+0.05}_{-0.03}$ & $8.38^{+4.92}_{-3.69}$ & 100 & $9.21^{+2.37}_{-2.27}$ & 63/60 \\
& & & & & & \\
\multicolumn{8}{l}{GRB 060428B (T+252s to T+418s)} \\
1   & $2.64^{+0.14}_{-0.14}$  & ...                    & ...                    & ...                    & ... & $2.28^{+1.63}_{-1.50}$ & 58/64  \\
2   & $2.58^{+0.15}_{-0.14}$  & $0.69^{+0.02}_{-0.03}$ & $<$ Inst. res.         & $0.62^{+0.37}_{-0.37}$ & 34 & $1.73^{+1.64}_{-1.51}$ & 50/62  \\
3   & $2.33^{+0.14}_{-0.22}$  & $< 1.07$               & $0.33^{+0.06}_{-0.17}$ & $9.71^{+5.48}_{-8.19}$ & 2100 & $0.39^{+1.93}_{-0.17}$ & 48/61 \\
\enddata
\tablenotetext{a}{Models: [1] Absorbed power law, [2] absorbed power law plus a narrow Gaussian (width restricted to less than the instrumental resolution) and [3] absorbed power law plus a free-width Gaussian. Models containing blackbody components are not reported in this table as the fits were poorly constrained. All errors are quoted at 90.0\% confidence.}
\label{table:Spectral_fits}
\end{deluxetable}
% ============================

% ============================
\begin{deluxetable}{crrrrrrrr}
\tablecaption{Summary of spectral fits for GRB 050822 (T+471 s to T+661 s).}
\tabletypesize{\footnotesize}
\tablecolumns{9}
\tablehead{
\colhead{Model$^{a}$} &
\colhead{Photon} &
\colhead{High energy} &
\colhead{Line energy} & 
\colhead{Line width} &
\colhead{Line norm.} &
\colhead{Equiv. width} &
\colhead{$N_{\rm H}$} &
\colhead{$\chi^{2}/\nu$} 
\\
\colhead{} &
\colhead{index} &
\colhead{cutoff} &
\colhead{(keV)} &
\colhead{(keV)} &
\colhead{($\times10^{-2}$ photons} &
\colhead{(eV)} &
\colhead{($\times10^{20}$} &
\colhead{} 
\\
\colhead{} &
\colhead{} &
\colhead{(keV)} &
\colhead{} &
\colhead{} &
\colhead{cm$^{-2}$ s$^{-1}$)} &
\colhead{} &
\colhead{cm$^{-2}$)} &
\colhead{} 
\\
}
\startdata
1 & $5.20^{+0.52}_{-0.44}$ & ...                    & ...                    & ...                    & ...                    & ...  & $34.0^{+7.7}_{-6.4}$ & 90/40 \\
2 & $2.84^{+0.57}_{-0.18}$ & ...                    & $< 0.47$               & $0.40^{+0.04}_{-0.07}$ & $49^{+10}_{-20}$       & 240 & $< 12.2$             & 49/37 \\
3 & $2.27^{+0.57}_{-0.18}$ & $0.51^{+0.01}_{-0.01}$ & ...                    & ...                    & ...                    & ...  & $18.0^{+7.1}_{-4.0}$ & 70/39 \\
4 & $2.25^{+0.12}_{-0.12}$ & $0.82^{+0.03}_{-0.03}$ & $0.68^{+0.04}_{-0.04}$ & $0.20^{+0.05}_{-0.04}$ & $8.50^{+1.64}_{-1.64}$ & 300 & $9.51^{+0.86}_{-0.79}$ & 55/36 \\
\enddata
\tablenotetext{a}{Models: [1] Absorbed power law, [2] absorbed power law plus a free-width Gaussian, [3] absorbed cutoff power law, [4] absorbed cutoff power law plus a free-width Gaussian.}
\label{table:050822}
\end{deluxetable}
% ============================

% ============================
% ============================

% ============================
% ============================

% ============================
% ============================

% ============================
% ============================

% ============================
% ============================

% ============================
% ============================

% ============================
% ============================

% ============================
% ============================

% ============================
% ============================

% ============================
% ============================

% ============================
% ============================

% ============================
% ============================

% ============================
% ============================

% ============================
% ============================

% ============================
% ============================

% ============================
% ============================

\clearpage

% ============================
% ============================

% ============================
% ============================


%% in parentheses, after the keyword, but they will not be verified.
\clearpage 

{\it Facilities:} \facility{Swift}.



\bibliographystyle{apj}
\bibliography{refs}

\begin{thebibliography}{70}
\expandafter\ifx\csname natexlab\endcsname\relax\def\natexlab#1{#1}\fi

\bibitem[{{Amati} {et~al.}(2000){Amati}, {Frontera}, {Vietri}, {in't Zand},
  {Soffitta}, {Costa}, {Del Sordo}, {Pian}, {Piro}, {Antonelli}, {Fiume},
  {Feroci}, {Gandolfi}, {Guidorzi}, {Heise}, {Kuulkers}, {Masetti},
  {Montanari}, {Nicastro}, {Orlandini}, \& {Palazzi}}]{Amati_2000}
{Amati}, L., {Frontera}, F., {Vietri}, M., {et~al.} 2000, Science, 290, 953

\bibitem[{{Antonelli} {et~al.}(2000){Antonelli}, {Piro}, {Vietri}, {Costa},
  {Soffitta}, {Feroci}, {Amati}, {Frontera}, {Pian}, {Zand}, {Heise},
  {Kuulkers}, {Nicastro}, {Butler}, {Stella}, \& {Perola}}]{Antonelli_2000}
{Antonelli}, L.~A., {Piro}, L., {Vietri}, M., {et~al.} 2000, \apjl, 545, L39

\bibitem[{{Arnaud}(1996)}]{Arnaud_1996}
{Arnaud}, K.~A. 1996, in ASP Conf. Ser. 101: Astronomical Data Analysis
  Software and Systems V, ed. G.~H. {Jacoby} \& J.~{Barnes}, 17--+

\bibitem[{{Ballantyne} \& {Ramirez-Ruiz}(2001)}]{Ballantyne_2001}
{Ballantyne}, D.~R. \& {Ramirez-Ruiz}, E. 2001, \apjl, 559, L83

\bibitem[{{Brainerd} {et~al.}(1994){Brainerd}, {Paciesas}, {Meegan}, \&
  {Fishman}}]{Brainerd_1994}
{Brainerd}, J.~J., {Paciesas}, W.~S., {Meegan}, C.~A., \& {Fishman}, G.~J.
  1994, in AIP Conf. Proc. 307: Gamma-Ray Bursts, ed. G.~J. {Fishman}, 122

\bibitem[{{Burrows} {et~al.}(2005{\natexlab{a}}){Burrows}, {Hill}, {Nousek},
  {Kennea}, {Wells}, {Osborne}, {Abbey}, {Beardmore}, {Mukerjee}, {Short},
  {Chincarini}, {Campana}, {Citterio}, {Moretti}, {Pagani}, {Tagliaferri},
  {Giommi}, {Capalbi}, {Tamburelli}, {Angelini}, {Cusumano}, {Br{\"a}uninger},
  {Burkert}, \& {Hartner}}]{Burrows_2005}
{Burrows}, D.~N., {Hill}, J.~E., {Nousek}, J.~A., {et~al.} 2005{\natexlab{a}},
  Space Science Reviews, 120, 165

\bibitem[{{Burrows} {et~al.}(2005{\natexlab{b}}){Burrows}, {Romano}, {Falcone},
  {Kobayashi}, {Zhang}, {Moretti}, {O'Brien}, {Goad}, {Campana}, {Page},
  {Angelini}, {Barthelmy}, {Beardmore}, {Capalbi}, {Chincarini}, {Cummings},
  {Cusumano}, {Fox}, {Giommi}, {Hill}, {Kennea}, {Krimm}, {Mangano},
  {Marshall}, {M{\'e}sz{\'a}ros}, {Morris}, {Nousek}, {Osborne}, {Pagani},
  {Perri}, {Tagliaferri}, {Wells}, {Woosley}, \& {Gehrels}}]{Burrows_2005Sci}
{Burrows}, D.~N., {Romano}, P., {Falcone}, A., {et~al.} 2005{\natexlab{b}},
  Science, 309, 1833

\bibitem[{{Butler} {et~al.}(2005){Butler}, {Ricker}, {Vanderspek}, {Ford},
  {Crew}, {Lamb}, \& {Jernigan}}]{Butler_2005}
{Butler}, N., {Ricker}, G., {Vanderspek}, R., {et~al.} 2005, \apjl, 627, L9

\bibitem[{{Butler}(2007)}]{Butler_2007}
{Butler}, N.~R. 2007, \apj, 656, 1001

\bibitem[{{Butler} \& {Kocevski}(2007)}]{Butler_Kocevski_2007}
{Butler}, N.~R. \& {Kocevski}, D. 2007, \apj, 663, 407

\bibitem[{{Campana} {et~al.}(2006{\natexlab{a}}){Campana}, {Beardmore},
  {Cusumano}, \& {Godet}}]{Campana_2006_caldb}
{Campana}, S., {Beardmore}, A.~P., {Cusumano}, G., \& {Godet}, .
  2006{\natexlab{a}}, {\it Swift\/} XRT CALDB Release Notes, 1

\bibitem[{{Campana} {et~al.}(2006{\natexlab{b}}){Campana}, {Mangano},
  {Blustin}, {Brown}, {Burrows}, {Chincarini}, {Cummings}, {Cusumano}, {Della
  Valle}, {Malesani}, {Meszaros}, {Nousek}, {Page}, {Sakamoto}, {Waxman},
  {Zhang}, {Dai}, {Gehrels}, {Immler}, {Marshall}, {Mason}, {Moretti},
  {O'Brien}, {Osborne}, {Page}, {Romano}, {Roming}, {Tagliaferri}, {Cominsky},
  {Giommi}, {Godet}, {Kennea}, {Krimm}, {Angelina}, {Barthelmy}, {Boyd},
  {Palmer}, {Wells}, \& {White}}]{Campana_2006}
{Campana}, S., {Mangano}, V., {Blustin}, A.~J., {et~al.} 2006{\natexlab{b}},
  \nat, 442, 1008

\bibitem[{{Campana} {et~al.}(2006{\natexlab{c}}){Campana}, {Romano}, {Covino},
  {Lazzati}, {de Luca}, {Chincarini}, {Moretti}, {Tagliaferri}, {Cusumano},
  {Giommi}, {Mangano}, {Perri}, {La Parola}, {Capalbi}, {Mineo}, {Antonelli},
  {Burrows}, {Hill}, {Racusin}, {Kennea}, {Morris}, {Pagani}, {Nousek},
  {Osborne}, {Goad}, {Page}, {Beardmore}, {Godet}, {O'Brien}, {Wells},
  {Angelini}, \& {Gehrels}}]{Campana_2006_AA}
{Campana}, S., {Romano}, P., {Covino}, S., {et~al.} 2006{\natexlab{c}}, \aap,
  449, 61

\bibitem[{{Cenko} {et~al.}(2006){Cenko}, {Berger}, \& {Cohen}}]{GCN_4592}
{Cenko}, S.~B., {Berger}, E., \& {Cohen}, J. 2006, GRB Coordinates Network,
  4592, 1

\bibitem[{{Chen} {et~al.}(2005){Chen}, {Thompson}, {Prochaska}, \&
  {Bloom}}]{GCN3709}
{Chen}, H.-W., {Thompson}, I., {Prochaska}, J.~X., \& {Bloom}, J. 2005, GRB
  Coordinates Network Circular, 3709, 1

\bibitem[{{Chen} {et~al.}(2003){Chen}, {Ikebe}, \& {B{\"o}hringer}}]{Chen_2003}
{Chen}, Y., {Ikebe}, Y., \& {B{\"o}hringer}, H. 2003, \aap, 407, 41

\bibitem[{{Cucchiara} {et~al.}(2006){Cucchiara}, {Fox}, \& {Berger}}]{GCN_4729}
{Cucchiara}, A., {Fox}, D.~B., \& {Berger}, E. 2006, GRB Coordinates Network
  Circular, 4729, 1

\bibitem[{{De Grandi} \& {Molendi}(1999)}]{DeGrandi_1999}
{De Grandi}, S. \& {Molendi}, S. 1999, \aap, 351, L45

\bibitem[{{Dickey} \& {Lockman}(1990)}]{Dickey_1990}
{Dickey}, J.~M. \& {Lockman}, F.~J. 1990, \araa, 28, 215

\bibitem[{{Dupree} {et~al.}(2006){Dupree}, {Falco}, {Prochaska}, {Chen}, \&
  {Bloom}}]{GCN_4969}
{Dupree}, A.~K., {Falco}, E., {Prochaska}, J.~X., {Chen}, H.-W., \& {Bloom},
  J.~S. 2006, GRB Coordinates Network Circular, 4969, 1

\bibitem[{{Eadie} {et~al.}(1971){Eadie}, {Drijard}, {James}, {Roos}, \&
  B.}]{eadie}
{Eadie}, W.~T., {Drijard}, D., {James}, F.~E., {Roos}, M., \& B., S. 1971,
  {Statistical methods in experimental physics} (Amsterdam: North-Holland,
  1971)

\bibitem[{{Freeman} {et~al.}(1999){Freeman}, {Graziani}, {Lamb}, {Loredo},
  {Fenimore}, {Murakami}, \& {Yoshida}}]{Freeman_1999}
{Freeman}, P.~E., {Graziani}, C., {Lamb}, D.~Q., {et~al.} 1999, \apj, 524, 753

\bibitem[{{Frontera} {et~al.}(2004){Frontera}, {Amati}, {Zand}, {Lazzati},
  {K{\"o}nigl}, {Vietri}, {Costa}, {Feroci}, {Guidorzi}, {Montanari},
  {Orlandini}, {Pian}, \& {Piro}}]{Frontera_2004}
{Frontera}, F., {Amati}, L., {Zand}, J.~J.~M.~i., {et~al.} 2004, \apj, 616,
  1078

\bibitem[{{Galama} {et~al.}(1998){Galama}, {Vreeswijk}, {van Paradijs},
  {Kouveliotou}, {Augusteijn}, {Bohnhardt}, {Brewer}, {Doublier}, {Gonzalez},
  {Leibundgut}, {Lidman}, {Hainaut}, {Patat}, {Heise}, {in 't Zand}, {Hurley},
  {Groot}, {Strom}, {Mazzali}, {Iwamoto}, {Nomoto}, {Umeda}, {Nakamura},
  {Young}, {Suzuki}, {Shigeyama}, {Koshut}, {Kippen}, {Robinson}, {de Wildt},
  {Wijers}, {Tanvir}, {Greiner}, {Pian}, {Palazzi}, {Frontera}, {Masetti},
  {Nicastro}, {Feroci}, {Costa}, {Piro}, {Peterson}, {Tinney}, {Boyle},
  {Cannon}, {Stathakis}, {Sadler}, {Begam}, \& {Ianna}}]{Galama_1998}
{Galama}, T.~J., {Vreeswijk}, P.~M., {van Paradijs}, J., {et~al.} 1998, \nat,
  395, 670

\bibitem[{{Gehrels} {et~al.}(2004){Gehrels}, {Chincarini}, {Giommi}, {Mason},
  {Nousek}, {Wells}, {White}, {Barthelmy}, {Burrows}, {Cominsky}, {Hurley},
  {Marshall}, {M{\'e}sz{\'a}ros}, {Roming}, {Angelini}, {Barbier}, {Belloni},
  {Campana}, {Caraveo}, {Chester}, {Citterio}, {Cline}, {Cropper}, {Cummings},
  {Dean}, {Feigelson}, {Fenimore}, {Frail}, {Fruchter}, {Garmire}, {Gendreau},
  {Ghisellini}, {Greiner}, {Hill}, {Hunsberger}, {Krimm}, {Kulkarni}, {Kumar},
  {Lebrun}, {Lloyd-Ronning}, {Markwardt}, {Mattson}, {Mushotzky}, {Norris},
  {Osborne}, {Paczynski}, {Palmer}, {Park}, {Parsons}, {Paul}, {Rees},
  {Reynolds}, {Rhoads}, {Sasseen}, {Schaefer}, {Short}, {Smale}, {Smith},
  {Stella}, {Tagliaferri}, {Takahashi}, {Tashiro}, {Townsley}, {Tueller},
  {Turner}, {Vietri}, {Voges}, {Ward}, {Willingale}, {Zerbi}, \&
  {Zhang}}]{Gehrels_2004}
{Gehrels}, N., {Chincarini}, G., {Giommi}, P., {et~al.} 2004, \apj, 611, 1005

\bibitem[{{Gelman} {et~al.}(1995){Gelman}, {Carlin}, {Stern}, \& B.}]{gelman}
{Gelman}, A., {Carlin}, J.~B., {Stern}, H.~S., \& B., R.~D. 1995, Bayesian Data
  Analysis (London: Chapman \& Hall)

\bibitem[{{Gelman} {et~al.}(1996){Gelman}, {Meng}, \& {Stern}}]{gelman_1996}
{Gelman}, A., {Meng}, X.-L., \& {Stern}, H.~S. 1996, Statistica Sinica, 6, 733

\bibitem[{{Ghisellini} \& {Celotti}(1999)}]{Ghisellini_1999}
{Ghisellini}, G. \& {Celotti}, A. 1999, \apjl, 511, L93

\bibitem[{{Gregory}(2005)}]{gregory2005}
{Gregory}, P.~C. 2005, {Bayesian Logical Data Analysis for the Physical
  Sciences} (Cambridge, UK: Cambridge University Press)

\bibitem[{{Hjorth} {et~al.}(2003){Hjorth}, {Sollerman}, {M{\o}ller}, {Fynbo},
  {Woosley}, {Kouveliotou}, {Tanvir}, {Greiner}, {Andersen}, {Castro-Tirado},
  {Castro Cer{\'o}n}, {Fruchter}, {Gorosabel}, {Jakobsson}, {Kaper}, {Klose},
  {Masetti}, {Pedersen}, {Pedersen}, {Pian}, {Palazzi}, {Rhoads}, {Rol}, {van
  den Heuvel}, {Vreeswijk}, {Watson}, \& {Wijers}}]{Hjorth_2003}
{Hjorth}, J., {Sollerman}, J., {M{\o}ller}, P., {et~al.} 2003, \nat, 423, 847

\bibitem[{{Holman} {et~al.}(2005){Holman}, {Garnavich}, \& {Stanek}}]{GCN3716}
{Holman}, M., {Garnavich}, P., \& {Stanek}, K.~Z. 2005, GRB Coordinates Network
  Circular, 3716, 1

\bibitem[{{Kass} \& {Raftery}(1995)}]{Kass}
{Kass}, R.~E. \& {Raftery}, A.~E. 1995, J. Am. Stat. Ass., 90, 773

\bibitem[{{Lazzati} {et~al.}(2004){Lazzati}, {Rossi}, {Ghisellini}, \&
  {Rees}}]{Lazzati_2004}
{Lazzati}, D., {Rossi}, E., {Ghisellini}, G., \& {Rees}, M.~J. 2004, \mnras,
  347, L1

\bibitem[{{Lee}(1989)}]{Lee_1989}
{Lee}, P.~M. 1989, {Bayesian Statistics: An Introduction} (New York: Halsted
  Press)

\bibitem[{{Loredo}(1990)}]{Loredo_1990}
{Loredo}, T.~J. 1990, in Maximum-Entropy and Bayesian Methods, Dartmouth., ed.
  P.~Fougere (Dordrecht, The Netherlands: Kluwer Academic Publishers), 81--142

\bibitem[{{Loredo}(1992)}]{Loredo_1992}
---. 1992, in Statistical Challenges in Modern Astronomy, Springer-Verlag., ed.
  D.~Feigelson \& G.~Babu (New York: Springer-Verlag), 275--297

\bibitem[{{Mewe} {et~al.}(1985){Mewe}, {Gronenschild}, \& {van den
  Oord}}]{Mewe_1985}
{Mewe}, R., {Gronenschild}, E.~H.~B.~M., \& {van den Oord}, G.~H.~J. 1985,
  \aaps, 62, 197

\bibitem[{{Mirabal} \& {Halpern}(2006)}]{GCN_4591}
{Mirabal}, N. \& {Halpern}, J.~P. 2006, GRB Coordinates Network, 4591, 1

\bibitem[{{Morrison} \& {McCammon}(1983)}]{Morrison_1983}
{Morrison}, R. \& {McCammon}, D. 1983, \apj, 270, 119

\bibitem[{{Pandey} {et~al.}(2006){Pandey}, {Castro-Tirado}, {McBreen},
  {P{\'e}rez-Ram{\'{\i}}rez}, {Bremer}, {Guerrero}, {Sota}, {Cobb},
  {Jel{\'{\i}}nek}, {de Ugarte Postigo}, {Gorosabel}, {Guziy}, {Guidorzi},
  {Bailyn}, {Mu{\~n}oz-Darias}, {Gomboc}, {Monfardini}, {Mundell}, {Tanvir},
  {Levan}, {Bhatt}, {Sahu}, {Sharma}, {Bogdanov}, \& {Combi}}]{Pandey_2006}
{Pandey}, S.~B., {Castro-Tirado}, A.~J., {McBreen}, S., {et~al.} 2006, \aap,
  460, 415

\bibitem[{{Pian} {et~al.}(2006){Pian}, {Mazzali}, {Masetti}, {Ferrero},
  {Klose}, {Palazzi}, {Ramirez-Ruiz}, {Woosley}, {Kouveliotou}, {Deng},
  {Filippenko}, {Foley}, {Fynbo}, {Kann}, {Li}, {Hjorth}, {Nomoto}, {Patat},
  {Sauer}, {Sollerman}, {Vreeswijk}, {Guenther}, {Levan}, {O'Brien}, {Tanvir},
  {Wijers}, {Dumas}, {Hainaut}, {Wong}, {Baade}, {Wang}, {Amati}, {Cappellaro},
  {Castro-Tirado}, {Ellison}, {Frontera}, {Fruchter}, {Greiner}, {Kawabata},
  {Ledoux}, {Maeda}, {M{\o}ller}, {Nicastro}, {Rol}, \& {Starling}}]{Pian_2006}
{Pian}, E., {Mazzali}, P.~A., {Masetti}, N., {et~al.} 2006, \nat, 442, 1011

\bibitem[{{Piran}(2005)}]{Piran_2005}
{Piran}, T. 2005, Reviews of Modern Physics, 76, 1143

\bibitem[{{Piranomonte} {et~al.}(2006){Piranomonte}, {D'Elia}, {Fiore},
  {Covino}, {Fugazza}, {Chincarini}, {Malesani}, {D'Avanzo}, {Antonelli},
  {Ledoux}, {Lopez}, \& {Naef}}]{GCN_4520}
{Piranomonte}, S., {D'Elia}, V., {Fiore}, F., {et~al.} 2006, GRB Coordinates
  Network Circular, 4520, 1

\bibitem[{{Piro} {et~al.}(1999){Piro}, {Costa}, {Feroci}, {Frontera}, {Amati},
  {dal Fiume}, {Antonelli}, {Heise}, {in 't Zand}, {Owens}, {Parmar},
  {Cusumano}, {Vietri}, \& {Perola}}]{Piro_1999}
{Piro}, L., {Costa}, E., {Feroci}, M., {et~al.} 1999, \apjl, 514, L73

\bibitem[{{Piro} {et~al.}(2000){Piro}, {Garmire}, {Garcia}, {Stratta}, {Costa},
  {Feroci}, {M{\'e}sz{\'a}ros}, {Vietri}, {Bradt}, {Frail}, {Frontera},
  {Halpern}, {Heise}, {Hurley}, {Kawai}, {Kippen}, {Marshall}, {Murakami},
  {Sokolov}, {Takeshima}, \& {Yoshida}}]{Piro_2000}
{Piro}, L., {Garmire}, G., {Garcia}, M., {et~al.} 2000, Science, 290, 955

\bibitem[{{Press} {et~al.}(1992){Press}, {Teukolsky}, {Vetterling}, \&
  {Flannery}}]{Press}
{Press}, W.~H., {Teukolsky}, S.~A., {Vetterling}, W.~T., \& {Flannery}, B.~P.
  1992, {Numerical recipes in FORTRAN. The art of scientific computing}
  (Cambridge: University Press, |c1992, 2nd ed.)

\bibitem[{{Prochaska} {et~al.}(2005){Prochaska}, {Chen}, {Bloom}, {O'Meara},
  {Burles}, \& {Thompson}}]{GCN3732}
{Prochaska}, J.~X., {Chen}, H.-W., {Bloom}, J.~S., {et~al.} 2005, GRB
  Coordinates Network Circular, 3732, 1

\bibitem[{{Protassov} {et~al.}(2002){Protassov}, {van Dyk}, {Connors},
  {Kashyap}, \& {Siemiginowska}}]{Protassov_2002}
{Protassov}, R., {van Dyk}, D.~A., {Connors}, A., {Kashyap}, V.~L., \&
  {Siemiginowska}, A. 2002, \apj, 571, 545

\bibitem[{{Rees} \& {M{\'e}sz{\'a}ros}(2000)}]{Rees_2000}
{Rees}, M.~J. \& {M{\'e}sz{\'a}ros}, P. 2000, \apjl, 545, L73

\bibitem[{{Reeves} {et~al.}(2002){Reeves}, {Watson}, {Osborne}, {Pounds},
  {O'Brien}, {Short}, {Turner}, {Watson}, {Mason}, {Ehle}, \&
  {Schartel}}]{Reeves_2002}
{Reeves}, J.~N., {Watson}, D., {Osborne}, J.~P., {et~al.} 2002, \nat, 416, 512

\bibitem[{{Reichart} \& {Price}(2002)}]{Reichart_2002}
{Reichart}, D.~E. \& {Price}, P.~A. 2002, \apj, 565, 174

\bibitem[{{Romano} {et~al.}(2006){Romano}, {Campana}, {Chincarini}, {Cummings},
  {Cusumano}, {Holland}, {Mangano}, {Mineo}, {Page}, {Pal'Shin}, {Rol},
  {Sakamoto}, {Zhang}, {Aptekar}, {Barbier}, {Barthelmy}, {Beardmore}, {Boyd},
  {Burrows}, {Capalbi}, {Fenimore}, {Frederiks}, {Gehrels}, {Giommi}, {Goad},
  {Godet}, {Golenetskii}, {Guetta}, {Kennea}, {La Parola}, {Malesani},
  {Marshall}, {Moretti}, {Nousek}, {O'Brien}, {Osborne}, {Perri}, \&
  {Tagliaferri}}]{Romano_2006}
{Romano}, P., {Campana}, S., {Chincarini}, G., {et~al.} 2006, \aap, 456, 917

\bibitem[{{Rutledge} \& {Sako}(2003)}]{Rutledge_2003}
{Rutledge}, R.~E. \& {Sako}, M. 2003, \mnras, 339, 600

\bibitem[{{Sako} {et~al.}(2005){Sako}, {Harrison}, \& {Rutledge}}]{Sako_2005}
{Sako}, M., {Harrison}, F.~A., \& {Rutledge}, R.~E. 2005, \apj, 623, 973

\bibitem[{{Schwartz}(1978)}]{schwartz_1978}
{Schwartz}, G. 1978, Ann. Stat., 6, 461

\bibitem[{{Shaviv} \& {Dar}(1995)}]{Shaviv_1995}
{Shaviv}, N.~J. \& {Dar}, A. 1995, \apj, 447, 863

\bibitem[{{Shemi}(1994)}]{Shemi_1994}
{Shemi}, A. 1994, \mnras, 269, 1112

\bibitem[{{Sivia}(1996)}]{Sivia_1996}
{Sivia}, D.~S. 1996, Oxford Univ. Press (Oxford)

\bibitem[{{Stanek} {et~al.}(2003){Stanek}, {Matheson}, {Garnavich}, {Martini},
  {Berlind}, {Caldwell}, {Challis}, {Brown}, {Schild}, {Krisciunas}, {Calkins},
  {Lee}, {Hathi}, {Jansen}, {Windhorst}, {Echevarria}, {Eisenstein}, {Pindor},
  {Olszewski}, {Harding}, {Holland}, \& {Bersier}}]{Stanek_2003}
{Stanek}, K.~Z., {Matheson}, T., {Garnavich}, P.~M., {et~al.} 2003, \apjl, 591,
  L17

\bibitem[{{Starling} {et~al.}(2005){Starling}, {Vreeswijk}, {Ellison}, {Rol},
  {Wiersema}, {Levan}, {Tanvir}, {Wijers}, {Tadhunter}, {Zaurin}, {Gonzalez
  Delgado}, \& {Kouveliotou}}]{Starling_2005}
{Starling}, R.~L.~C., {Vreeswijk}, P.~M., {Ellison}, S.~L., {et~al.} 2005,
  \aap, 442, L21

\bibitem[{{Tavecchio} {et~al.}(2004){Tavecchio}, {Ghisellini}, \&
  {Lazzati}}]{Tavecchio_2004}
{Tavecchio}, F., {Ghisellini}, G., \& {Lazzati}, D. 2004, \aap, 415, 443

\bibitem[{{Tyler} {et~al.}(2006){Tyler}, {Page}, {Goad}, \&
  {Osborne}}]{Tyler_2006}
{Tyler}, L., {Page}, K., {Goad}, M., \& {Osborne}, J. 2006, in ASP Conf. Ser.
  351: Astronomical Data Analysis Software and Systems XV, ed. C.~{Gabriel},
  C.~{Arviset}, D.~{Ponz}, \& S.~{Enrique}, 97

\bibitem[{{van Dyk} {et~al.}(2001){van Dyk}, {Connors}, {Kashyap}, \&
  {Siemiginowska}}]{vandyk_2001}
{van Dyk}, D.~A., {Connors}, A., {Kashyap}, V.~L., \& {Siemiginowska}, A. 2001,
  \apj, 548, 224

\bibitem[{{Vietri} {et~al.}(2001){Vietri}, {Ghisellini}, {Lazzati}, {Fiore}, \&
  {Stella}}]{Vietri_2001}
{Vietri}, M., {Ghisellini}, G., {Lazzati}, D., {Fiore}, F., \& {Stella}, L.
  2001, \apjl, 550, L43

\bibitem[{{Vreeswijk} \& {Jaunsen}(2006)}]{GCN_4974}
{Vreeswijk}, P. \& {Jaunsen}, A. 2006, GRB Coordinates Network Circular, 4974,
  1

\bibitem[{{Watson} {et~al.}(2003){Watson}, {Reeves}, {Hjorth}, {Jakobsson}, \&
  {Pedersen}}]{Watson_2003}
{Watson}, D., {Reeves}, J.~N., {Hjorth}, J., {Jakobsson}, P., \& {Pedersen}, K.
  2003, \apjl, 595, L29

\bibitem[{{Watson} {et~al.}(2002){Watson}, {Reeves}, {Osborne}, {O'Brien},
  {Pounds}, {Tedds}, {Santos-Lle{\'o}}, \& {Ehle}}]{Watson_2002}
{Watson}, D., {Reeves}, J.~N., {Osborne}, J., {et~al.} 2002, \aap, 393, L1

\bibitem[{{Waxman}(1997)}]{Waxman_1997}
{Waxman}, E. 1997, \apjl, 485, L5

\bibitem[{{Yoshida} {et~al.}(1999){Yoshida}, {Namiki}, {Otani}, {Kawai},
  {Murakami}, {Ueda}, {Shibata}, \& {Uno}}]{Yoshida_1999}
{Yoshida}, A., {Namiki}, M., {Otani}, C., {et~al.} 1999, \aaps, 138, 433

\bibitem[{{Zhang}(2007)}]{Zhang_2007}
{Zhang}, B. 2007, Chinese Journal of Astronomy and Astrophysics, 7, 1

\end{thebibliography}

\end{document}